\documentclass[10pt]{article} 
\usepackage{amssymb,amsmath}     

\begin{document}


\def\inta{\int_{-a}^a}
\def\Pl#1{\frac\pl {#1\pl#1}}


\def\sct#1{\hspace{0in}\\{$\scriptstyle(#1)$\vspace{-10pt}}}
\def\st{\scriptstyle}
\def\sst{\scriptscriptstyle}

\def\sl{\slshape}
\def\up{\upshape}

\def\kn#1{{\kern -#1 true cm}}
\def\VE{\vfill\eject}
\def\mathphys{$ \ (M\cap\F)\cap\F\ \top{??}\app \ \f$}


\newtheorem{defn}{Definition}
\newtheorem{thm}{Theorem}
\newtheorem{prop}{Proposition}
\newtheorem{cor}{Corollary}

\newcommand{\ci}{\cite}
\newcommand{\lab}{\label}
\newcommand{\eq}{\eqref}
\def\bl{\renewcommand{\baselinestretch}} 
\def\cl{\centerline}
\def\bib#1{\bibitem[#1]{#1}}


\def\lp{\left(}
\def\rp{\right)}
\def\lb{\left[}
\def\rb{\right]}
\def\la{\langle\, }
\def\LA{\left\langle\, }
\def\ra{\,\rangle} 
\def\RA{\right\rangle\, }
\def\LB{\left\lbrace}
\def\RB{\right\rbrace}

\def\+#1#2{{#1+#2}}
\def\ola#1{\overleftarrow{#1}} 
\def\top#1#2{\smash{\mathop{\hbox to .2cm{$#2$}}\limits^{#1}}}

\def\0#1{{(#1)}}
\def\1#1{{\hat #1}}
\def\2#1{{\tilde #1}}
\def\3#1{{\boldsymbol#1}}
\def\4#1{{\mathbb#1}}
\def\5#1{{\cal#1}}
\def\6#1{_{\sst#1}}
\def\7#1{{\bar#1}}
\def\8{\infty}
\def\9#1{^{\sst#1}}
\def\/#1{{\bf#1}}
\def\;#1{{\breve#1}}

\def\bb#1{{\bf#1}}
\def\bh#1{{\boldsymbol{\hat{#1}}}}
\def\bt#1{{\boldsymbol{\tilde#1}}}
\def\ol#1{\overline{#1}}
\def\rh#1{{\bf{\hat{#1}}}}
\def\rt#1{{\bf{\tilde{#1}}}}
\def\v{\hskip.5pt\9\sharp} 
\def\w{\hskip.5pt\9\flat}
\def\wh#1{\widehat{#1}}    
\def\wt#1{\widetilde{#1}}


\def\a{\alpha} 
\def\b{\beta} 
\def\c{\chi}
\def\d{\delta} 
\def\e{\varepsilon} 
\def\f{\phi} 
\def\vf{\varphi} 
\def\g{\gamma}
\def\h{\eta} 
\def\i{\iota}
\def\k{\kappa}
\def\l{\lambda} 
\def\m{\mu} 
\def\n{\nu}
\def\o{\omega} 
\def\p{\pi} 
\def\vp{\varpi}
\def\q{\theta} 
\def\vq{\vartheta} 
\def\r{\rho}
\def\vr{\varrho} 
\def\R#1{{\rm #1}}
\def\s{{\sigma}} 
\def\vs{\varsigma}
\def\t{\tau} 
\def\u{\upsilon} 
\def\x{\xi}
\def\y{\psi} 
\def\z{\zeta}

\def\D{\Delta} 
\def\F{\Phi} 
\def\G{\Gamma}
\def\L{\Lambda} 
\def\O{\Omega} 
\def\P{\Pi}
\def\Q{\Theta} 
\def\S{\Sigma} 
\def\U{\Upsilon}
\def\Y{\Psi}
\def\X{\Xi}


\def\hb#1{{\qq\text{#1}\qq}}
\def\bull{$\bullet\ $}
\def\db{{d\kern-1.0ex {^-}}}
\def\diamond{$\diamondsuit\ $}
\def\heart{$\heartsuit\ $}
\def\spade{$\spadesuit\ $}
\def\club{$\clubsuit\ $}

\def\={\equiv} 
\def\app{\approx} 
\def\cc#1{{{\mathbb C\hskip.5pt}^{#1}}}
\def\ccc#1{{\mathbb C\hskip.5pt}^{#1+1}}
\def\clif#1{C\!\ell_{#1}\,} 
\def\cliff#1{C\!\ell_{#1+1}\,} 
\def\curl{\nabla\times} 
\def\ptl#1#2{{\frac{\partial #1}{\partial#2}}}
\def\div{\nabla\cdot } 
\def\eminus#1{e^{-2\p #1}\,}
\def\eplus#1{e^{2\p#1}\,} 
\def\grad{\nabla} 
\def\ie{{\sl i.e.,}}
\def\im{{\,\rm Im}\ }  
\def\imp{\ \Rightarrow\ }
\def\i1#1{\int_{-\infty}^\infty d#1\, } 
\def\inv{^{-1}}
\def\intt{\int\!\!\!\int}
\def\ir{\int_{-\infty}^\infty} 
\def\llr{L^2(\3R)}
\def\lra{\leftrightarrow}
\def\Lra{\Leftrightarrow} 
\def\lv#1{{\,\top{\leftarrow}#1\,}}
\def\nn{{n+1}}
\def\pl{\partial}
\def\qq{\quad} 
\def\qqq{\qquad} 
\def\qed{\vrule height6pt width4pt depth 0pt}
\def\re{{\,\rm Re}\  }   
\def\rr#1{{{\mathbb R}^{#1}}}
\def\rrr#1{{\4R^{#1+1}}}
\def\sgn{{\rm sgn \,}}
\def\sh#1{\hskip#1ex} 
\def\sr{\sqrt}
\def\supp{{\rm supp \,}}
\def\sv#1{\vskip#1ex}


\title{Complex-Distance Potential Theory and Hyperbolic
Equations }

\author{Gerald Kaiser\\
   Virginia Center for Signals and Waves\\
	kaiser@wavelets.com $\bullet$\ www.wavelets.com}

\maketitle                  

\begin{abstract}\noindent
An extension of potential theory in $\rr n$ is obtained by
continuing the Euclidean distance function holomorphically to
$\cc n$. The resulting  Newtonian potential is generated by an
extended source distribution $\2\d(\3z)$ in $\cc n$ whose
restriction to $\rr n$ is the point source $\d(\3x)$. This
provides a possible model for extended particles in physics. In
$\ccc n$, interpreted as complex \sl spacetime, \rm $\2\d$
acts as a \sl propagator \rm  generating solutions of the wave
equation from their initial values. This gives a new
connection between elliptic and hyperbolic equations that
does not assume analyticity of the Cauchy data. Generalized
to Clifford analysis, it induces a similar connection between
solutions of elliptic and hyperbolic Dirac equations. There is
a natural application to the time-dependent, inhomogeneous
Dirac and Maxwell equations, and the `electromagnetic
wavelets'  introduced previously are an example.

\end{abstract}

\section{Motivation and Preliminaries}
\label{motivation}

Most fundamental theories of physics are based on the concept
of potentials and fields generated by   \sl point sources, \rm
which presupposes that objects or ``particles'' can,  in
principle, be localized within arbitrarily small regions of
space and/or time. This is a vast extrapolation from
empirical evidence, and it should perhaps not come as a
surprise if such theories experience some fundmental
difficulties. In Newtonian mechanics, the problem of $N$ 
point ``bodies'' interacting through gravitation has, in general,
no solution due to the possibility of collisions. This becomes
a serious difficulty for $N\ge 3$, where the set of initial
conditions leading to  collisions is nontrivial \ci[Chapter
10]{AM78}. In classical electrodynamics,   difficulties arise
where the field produced by a point charge unavoidably acts
back on the same charge, leading to infinite self-energies and
run-away particle trajectories \ci{J99}. A way out of this
dilemma was proposed by Wheeler and Feynman  \ci{WF45,
WF49}, but their
\sl action-at-a distance
\rm theory,  apart from being highly counter-intuitive
\ci[p.~28-8]{F64},  has resisted quantization and is not
generally regarded as being a fundamental description of
Nature. In quantum electrodynamics and other quantum field
theories, point particles cause divergences which necessitate
infinite ``renormalization'' procedures, a subject of some
contraversy \ci{C99}.  String theory \ci{P98} does, in fact, not
need infinite renormalization because its basic objects
(strings) are  extended in space rather than mathematical
points. This is one of the reasons it is regarded with great
hope as a possibility for unifying physical theories. However, a
full development of (super)string  physics is rather difficult
and not expected to near completion for many years. 

The ideas developed here began many years ago,
motivated in part by the hope that an extension of physics to
\sl complex spacetime, \rm justified at the
foundational level, might give a way to circumvent the
problems associated with point sources by applying residue
methods. After some years of study and research  this led to
papers \ci{K77,K78,K80,K87} and books \ci{K90,K94} whose
main thrust has been to develop a direct physical 
interpretation of the complex spacetime as an \sl extended
phase space, \rm  with the imaginary spacetime parameters
carrying directional information while the real spacetime
parameters describe (approximate!) localization.  A major
stumbling block in this program has been the construction of 
extended  sources, since it seemed that they tend to spoil the
holomorphy of the theory globally rather than just locally.
Here we propose a natural solution to this problem. It turns
out that although holomorphy enters the theory at the level of
the fundamental potential or Green's function  and this indeed
gives a canonical extension of general fields to complex
spacetime through convolution, these fields need not be
holomorphic \sl anywhere \rm due to the specific structure
of the extended sources. 

We begin in Section \ref{S: CxDist}  by  extending the
Euclidean distance $r(\3x)$ to a function
$\g(\3x+i\3y)$ holomorphic in a domain  $\5O_n\subset\cc n$.
This leads to a natural coordinate system in $\rr n$
which will play an important role in the sequel:  the  \sl
oblate spheroidal coordinates \rm adapted to $\3y\ne\30$.  
 
In Sections \ref{S: AnalPot} and \ref{S: SingSource}  we use
$\g(\3z)$ to extend the Newtonian potential $\f(\3x)$
(fundamental solution of the Laplacian in $\rr n$) to 
$\5O_n$. The resulting \sl holomorphic potential \rm
$\f(\3z)$ has a source distribution
$\2\d(\3z)$ in $\cc n$ which, although nowhere
holomorphic,  is nevertheless a canonical extension to $\cc n$
of the point source $\d(\3x)$ in $\rr n$.  For \sl even \rm
$n\ge 4$, $\2\d(\3x+i\3y)$ with fixed $\3y\ne\30$ is
supported in $\3x$ on the \sl sphere \rm $\5B(\3y)$ of
codimension 2 and radius $|\3y|$,  centered at the origin  and
lying in the hyperplane orthogonal to $\3y$.  For all other
values of $n$, it is supported in $\3x$ on a \sl membrane
\rm whose boundary is $\5B(\3y)$.  The membrane is 
determined by a choice of branch cut in $\g$, and in the
simplest case it is a \sl disk \rm $E\60(\3y)$.

In Section \ref{S: CompInR2R3} we compute the sources
$\2\d(\3x+i\3y)$ explicitly as distributions in $\rr3$ and
$\rr4$ supported, respectively, in $E\60(\3y)$ and
$\5B(\3y)$.  This is especially interesting in the
physical case of $\rr3$, where  $\2\d(\3x+i\3y)$ consists of
a uniform line charge on the rim of the disk $E\60(\3y)$
accompanied by  a simple layer (with zero net charge) and a
double layer distributed on the interior. Hence the original
unit ``charge'' carried by $\d(\3x)$ is now spread over a circle
of radius $|\3y|$, and a variable polarization density is
induced on the disk swept out by ``blowing up'' the point
source from the origin to its rim.  That so much natural
structure results from a simple analytic continuation is
remarkable.

In Section \ref{S: Waves} we show how to include \sl time
\rm in this framework. Since we already have a
holomorphic extension $\f(\3x, s+it)$  of the fundamental
solution $\f(\3x, s)$ of the Laplacian in $\rrr n$, it
is natural to interpret $t$ as a time parameter and look for a
connection with the wave operator
$\D_\3x-\pl_t^2$, which is the ``analytic continuation'' to
$\4R^{n,1}$ of the Laplacian
$\D_\3x+\pl_s^2$ in $\rr\nn$, now interpreted as Euclidean
spacetime. We show that $\2\d(\3x, s+it)$ 
acts as a \sl propagator, \rm generating solutions of the
initial-value problem for the wave equation in $\4R^{n,1}$
from Cauchy data at $t=0$.  This gives a  connection
between the Laplace equation and the wave equation 
without assuming analyticity of the initial data, because the
propagator $\2\d(\3x, s+it)$ is not holomorphic in $s+it$.

In Section \ref{S: Clifford} we extend our method to
Clifford analysis, where the Laplacian and the wave operator
are replaced by their ``square roots,'' the \sl elliptic and
hyperbolic Dirac operators. \rm The time-dependent Maxwell
equations are a natural application of the ensuing formalism.

The idea of extending harmonic functions to obtain solutions
of the wave equation has been studied by Garabedian \ci{G64}
and, in a more general context,  by Ryan \ci{R90, R90a, R96a}.
However,  the methods employed in these references depend
critically on the assumption that the boundary data for the
harmonic functions is holomorphic. As emphasized above, our
method is free of this restriction.

\section{Complex Distance and Spheroidal Coordinates}
\lab{S: CxDist}
For $n\ge 2$, let 
\begin{align} \label{z}
\3z=\3x+i\3y\in\cc n, \qqq
|\3x|=r,\  |\3y|=a.
\end{align} 
Define the \sl complex length \rm of  $\3z$ as
\begin{align} \label{zeta}
\g (\3z)
\=\sr{\3z^2}=\sr{\3x^2-\3y^2+2i\3x\cdot\3y}
=\sr{r^2-a^2+2i\3x\cdot\3y}\=p+iq. 
\end{align} 
The  \sl complex distance \rm between any two points
$\3z_1, \3z_2\in\cc n$ is then defined as
$\g (\3z_1-\3z_2)$.  To complete this definition,
a branch of the complex root must be chosen.  The branch
points form the \sl null cone \rm
\begin{align}\label{C}
\5N=\LB\3z\mid \g(\3z)=0\RB
=\LB\3x+i\3y\mid 
\3x^2=\3y^2 \text{ and } \3x\cdot\3y=0\RB,
\end{align} 
a manifold of real dimension $2n-2$ in $\cc n\app\4R^{2n}$.
In the context of the wave equation, $\5N$ will be seen to be
related to the \sl light cone. \rm 
We will be interested in the restriction of $\5N$ to a
fixed nonzero $\3y\in\rr n$, \it i.e., \rm 
\begin{align}\label{By}
\5B\=\5B(\3y)=\LB \3x\in\rr n\mid 
r=a,\   \3x\cdot\3y=0\RB.
\end{align}
$\5B$ is the \sl sphere  \rm of dimension $n-2$ in $\rr n$
obtained by intersecting the sphere $r=a$ with the hyperplane
orthogonal to $\3y$.  We call $\5B(\3y)$ the \sl branching
sphere with  axis vector \rm $\3y$.

We  now concentrate on $\g (\3x+i\3y)$ as a function of
$\3x$, regarding $\3y$ as a fixed nonzero vector. In order to
make this function single-valued, it is necessary to introduce
a branch cut in $\rr n$.  To see how this can be done, fix any 
$\3x\in \5B$. In the plane determined by the two orthogonal
vectors $\3x$ and
$\3y$,  draw a  circle of radius $\e\le a$ centered at
$\3x$. Each time we go around this circle, 
$\g $ changes sign. In order to obtain a single-valued
function for $\g $, it is therefore necessary to cut every such
circle. Furthermore, this must be done subject to the
requirement that $\g $ must be an extension to $\cc n$ of the
usual length function in $\rr n$, i.e., 
\begin{align}\label{zeta-to-r}
   \3y\to\30\imp \g(\3x+i\3y) \to r(\3x)\ge 0.
\end{align}
The simplest cut (but certinainly not the only one) is obtained
by requiring that
\begin{align}  \label{upositive}
p\=\re\g \ge 0,
\end{align}
 which means that each of the above circles is cut at the
point 
\begin{align*}
\3x\6C=(a-\e)\bh x,\qqq \bh x\=\3x/r.
\end{align*}
 The totality of such points, as $\3x$ varies over $\5B$
and $0<\e\le a$, together with  $\5B$ itself ($\e=0$),  forms
the set
\begin{align}\label{E0}
 E\60\=  E\60(\3y)
=\LB \3x\in\rr n\mid r\le a, \ \3x\cdot\3y=0\RB,
\end{align}
 which is the \sl disk \rm of dimension $n-1$ in $\rr n$
obtained by intersecting the ball $r\le a$ with the hyperplane
through the origin orthogonal to $\3y$.  The boundary or \sl
rim
\rm of
$ E\60$ is
$\5B$:
\begin{align*}
\pl  E\60=\5B.
\end{align*}
(For $n=3$, $ E\60$ is indeed the disk of radius $a$ orthogonal
to $\3y$ and $\5B$ is its rim.)  In the context of holomorphic
potential theory, $ E\60$ or $\5B$, depending on whether $n$
is odd or even, will represent the support of a source
distribution. As $\3y\to \30$,  both sets contract to the
origin,  the support of the usual point source.

We now derive the basic properties  of the real and imaginary
parts $p, q$ of $\g $, which will be important in our study of
the holomorphic potentials.  From  \eq{zeta} it follows that 
\begin{equation}
\begin{aligned}\label{p,q}
p^2-q^2=r^2-a^2, \qqq pq=\3x\cdot\3y\=a\z\,,
\end{aligned} 
\end{equation}
where $\z$ is the projection
of $\3x$ onto $\3y\ne \30$.  Define the 
cylindrical coordinate $\r$ by
\begin{align} \label{rho}
\r^2=r^2-\z^2=a^2+p^2-q^2-\frac{p^2q^2}{a^2}
=\frac{(a^2+p^2)(a^2-q^2)}{a^2}.
\end{align}
This shows that  the imaginary part of $\g $ is bounded by
$a=|\3y|$:
\begin{align}\label{vbound}
-a\le  q\le a.
\end{align}
We will use $(\r,\z)$ as part of a cylindrical coordinate
system in $\rr n$ with $\3y$ as its ``$z$-axis.''
Equations \eq{p,q} and \eq{rho}  show that the 
surface  $ E_p=\{ \3x\mid  p=\text{constant}\}$
with a given value of $p>0$  is an \sl oblate spheroid \rm
given by
\begin{align}\label{Ep}
 E_p\colon\qq \frac{\r^2}{a^2+p^2}+\frac{\z^2}{p^2}=1.
\end{align}
 Similarly, the level surface 
$\{\3x\mid  q^2=\text{constant}\}$ with a given value of
$ 0<q^2<a^2$  is a \sl hyperboloid of one sheet \rm given by
\begin{align}\label{Hq}
 H_q\colon\qq \frac{\r^2}{a^2-q^2}-\frac{\z^2}{q^2}=1.
\end{align}
Note that we have so far avoided the surfaces with $p=0$ and
$q=0,\pm a$.   These give degenerate forms
of $ E_p$ and $ H_q$.  In fact, as $p\to+0$, the oblate
spheroid $ E_p$ contracts to the disk \eq{E0}:
\begin{align}
p\to 0\imp  E_p\to \LB\3x\mid \z=0,\ \r\le a\RB=  E\60\,.
\end{align}
More precisely, since the interior of $  E\60$ is covered 
twice,  we will distinguish between its \sl front and back
sides: \rm 
\begin{align}\lab{decomp/E0} 
E\60 =  E\60\9+\cup  E\60\9-\cup \5B\,,
\end{align}
 where
\begin{align}\label{E0pm}
  E\60\9\pm=\LB\3x\mid \z=\pm 0,\ \r< a\RB
=\LB \3x\mid p=0,\ \pm q> 0 \RB
\end{align}
are the \sl interiors \rm of the front and back sides of
$E\60$.  Although $  E\60\9+$ and $  E\60\9-$ coincide as
sets, the distinction between them will be very important for
the following reason.  It can be easily seen that
\begin{align}\label{jump}
\3x\in  E\60\9\pm\imp
q=\pm\sr{a^2-r^2}=\pm\sr{a^2-\r^2}.
\end{align}
 This shows that while $p$ is continuous across $  E\60$, $q$
has a spherical jump discontinuity there.  In the context of
holomorphic potential theory, $  E\60\9+$ and $  E\60\9-$
will be seen to form the two sides of a \sl double layer. \rm 

The degenerate values of $q$ are $q=0, \pm a$.
As $q\to  \pm a$, the semi-hyperboloids 
$ H_q\9\pm$ with $\pm q>0$ collapse to the positive and
negative $\z$-axis, respectively. As $q\to 0$, $ H_q$
collapses to the set $H_0=\{\3x\mid \z=0,\ \r\ge a\}$
which, like $  E\60$,  is covered twice, once by the limit
of each  semi-hyperboloid $ H_q\9\pm$ as $q\to\pm0$. But in
this case  we do not distinguish between the two copies
because $\g $ is continuous across $ H_0$.

As mentioned above, the parameters $\z$ and $\r$  are part of
a  cylindrical coordinate system  in $\rr n$. Note that the
intersection of $ E_p$ and $ H_q$ is
\begin{align*}
S_\g &= E_p\cap  H_q 
=\LB\3x\in\rr n\mid  a\z =pq, \  
a\r =\sr{a^2+p^2}\sr{a^2-q^2}\RB.
\end{align*}
Here and henceforth, we denote the
pair of real variables $(p,q)$ by the single complex variable
$\g =p+iq$ for convenience.  If $q\ne \pm a$, then $S_\g $ is
a  sphere of dimension $n-2$ and radius $\r$.   The variable
point on $S_\g $ may be represented  as
\begin{align}
\3x=\r \,\3\s+\z\,\bh y\in S_\g\,,
\end{align}
where $\3\s$ runs over the unit sphere of
dimension $n-2$ in the hyperplane orthogonal to $\3y$.  In
particular, $\g =0$ gives $\z=0$ and $\r=a$, so that
\begin{align}\notag
S\60=\{a\,\3\s\mid \3\s\in S^{n-2}\}=\5B(\3y).
\end{align} 
If $n=2$,   $S_\g $ consists of just two points. If $n=3$, 
$S_\g $ is a circle. Thus, for fixed $\3y\ne\30$,  a complete
set of \sl cylindrical coordinates \rm in $\rr n$ is given by 
$(\r, \z, \3\s)$.  Since $\z$ and $\r$ are functions of $(p,q)$, 
we may equivalently use the coordinates 
\begin{align}\label{OS}
\3x=(p,q, \3\s)\=(\g , \3\s).
\end{align}
 This is the \sl oblate spheroidal  coordinate system
\rm in $\rr n$,  long known as a useful tool in the theory of
special functions  \ci{T96}.  It is remarkable that this
system is so intimately related to the holomorphic extension
of the distance function.   

Now let $\nabla$ be the gradient with respect
to $\3x$ with $\3y$ held constant, and let $\D=\nabla^2$ be
the Laplacian in $\rr n$. Then
\begin{align}\label{zetagradzeta}
&\g ^2=\3z^2\imp \g \,\nabla\g =\3z, \hb{hence}
(\nabla\g )^2\=\nabla\g \cdot\nabla\g  =1.
\end{align}
 Taking the divergence of 
\eq{zetagradzeta} gives
\begin{align} \label{D/zeta}
\nabla\g \cdot\nabla\g +\g \D\g =n,
\imp \D\g =\frac{n-1}\g .
\end{align}
Now
\begin{align}\label{gradzeta}
\nabla\g =\frac{\3z}\g =\frac{\7\g \,\3z}{\7\g \g }\imp
\nabla p=\frac{p\,\3x+q\,\3y}{\7\g \,\g }, \qq
\nabla q=\frac{p\,\3y-q\,\3x}{\7\g \,\g },
\end{align}
therefore
\begin{gather}\label{orth}
\nabla\g \cdot\nabla\g=1 \imp
(\nabla p)^2-(\nabla q)^2=1, \qq \nabla p\cdot \nabla q=0\\
\nabla\7\g \cdot\nabla\g =\frac{|\3z|^2}{\7\g \g }
\imp(\nabla p)^2+(\nabla q)^2=\frac{r^2+a^2}{\7\g \g },
\end{gather}
which gives
\begin{equation}\label{grad/p2q2}
\begin{align}
(\nabla p)^2=\frac{a^2+p^2}{\7\g \g }, \qqq
(\nabla q)^2=\frac{a^2-q^2}{\7\g \g }.
\end{align} 
\end{equation}
 Finally, the real and imaginary parts of \eq{D/zeta}
give
\begin{align}\label{Du/Dv}
\D p=\frac {n-1}{\7\g \,\g }\,p,\qqq 
\D q=-\frac {n-1}{\7\g \,\g }\,q\,. 
\end{align}
We will need to compute volume integrals in the 
oblate spheroidal coordinates. Note that the area of the unit
sphere $S^{n-1}\subset\rr n$ is \ci{CH62}
\begin{align}\label{area}
\o_n=\frac{2\p^{n/2}}{\G(n/2)},\qqq n\ge 2.
\end{align}
Let $d\3\s$ denote the surface
measure on the unit sphere $S^{n-2}$  in the hyperplane
orthogonal to $\3y\ne\30$, \sl normalized \rm so that
\begin{align}\label{dsigma}
\int_{S^{n-2}} d\3\s=1.
\end{align}
Then for fixed $\3y$,  the volume measure  $d\3x$ in $\rr n$
is given in cylindrical coordinates by
\begin{align}\label{measures}
d\3x=\o_{n-1}\,\r^{n-2} d\r\, d\z\,d\3\s. 
\end{align}
By \eq{p,q} and \eq{rho},
\begin{align*}
\r^{n-2} d\r\, d\z&=\frac{\r^{n-3}}2\,d[\r^2] \,d\z
=\frac{\r^{n-3}}{2a^3}
\,d\lb (a^2+p^2)(a^2-q^2)\rb  d\lb pq \rb\\
&=a\inv \,\r^{n-3}(p\,dp-q\,dq)(p\,dq+q\,dp)
=a\inv \,\r^{n-3}(p^2+q^2)\,dp \, dq,
\end{align*}
where $dp\, dq$ denotes the antisymmetric exterior
product of differential forms (see \ci{GS64}, for example).
Therefore the volume measure in the oblate spheroidal  
coordinates is given by
\begin{align}\label{vol/os}
d\3x=\frac{\o_{n-1}}a\,\r^{n-3}\,(p^2+q^2) \,dp\,dq\,d\3\s
=\frac{\o_{n-1}}a\,\r^{n-3}\,\7\g \g \,dp\,dq\,d\3\s.
\end{align}

\section{Holomorphic Potentials  and Their Sources}
\lab{S: AnalPot}

For simplicity, we now assume that $n\ge 3$.  The case $n=2$
is similar but requires some special attention and will
be described elsewhere. Consider the \sl fundamental solution
\rm $\f(\3x)$ of Laplace's equation defined by
\begin{align}\lab{fund/real}
\D \f(\3x)=\d(\3x), \qq \lim_{r\to\8}\f(\3x)=0.
\end{align}
It  is given uniquely by \ci{CH62} 
\begin{align}\lab{newton}
\f(\3x)=\frac1{\o_n}\,
\frac{r^{2-n}}{2-n}\,,\qqq \3x\in\rr n,\  n\ge 3.
\end{align}
For $n=3$,  $\f(\3x)=-1/4\p r$  is the \sl Newton-Coulomb
potential  \rm with unit mass or charge. Define the \sl
holomorphic potential \rm in $\cc n$ by
\begin{align}\label{phi}
\f(\3z)=\frac1{\o_n}\,
\frac{\g^{2-n}}{2-n}, \qqq \3z\in\cc n, \ n\ge 3.
\end{align}
For  odd $n$, $\f(\3z)$ inherits the branch
cut of $\g (\3z)$. For even $n$, the only
singularities occur on $\5B$, where $\g =0$. Thus $\g(\3z)$
and  $\f(\3z)$ are analytic continuations of $r(\3x)$ and
$\f(\3x)$  to the domains
\begin{align*}
\5O_n&=\{\3z\in\cc n \mid  p>0\}
=\LB\3x+i\3y\mid \3x\notin   E\60(\3y)\RB, 
&& \text{odd } n\ge3\\
\5O_n&=\{\3z\in\cc n\mid \g \ne 0\}
=\LB\3x+i\3y\mid  \3x\notin\5B(\3y) \RB, 
&&\text{even } n\ge4.
\end{align*}

\begin{prop}\label{T: harmonic}
For fixed $\3y$,
$\f(\3x+i\3y)$ is harmonic with respect
to $\3x$ when $\3x+i\3y\in\5O_n$.  
\end{prop}

\sl Proof: \rm By \eq{phi}, we have 
\begin{align*}
\o_n\nabla\f(\3z)&=\g ^{1-n}\,\nabla\g .
\end{align*}
Thus by 
\eq{D/zeta},
\begin{align*}
\o_n\D\f(\3z)&=(1-n)\g ^{-n}+\g ^{1-n}\D\g =0.\sh2\qed
\end{align*}
Our objective is to compute the \sl source
distribution \rm of $\f(\3z)$, which we define 
formally by analogy with \eq{fund/real} as
\begin{align}\label{delta}
\2\d(\3z)=\D\f(\3z), \qqq \3z\in\cc n.
\end{align}
 This will be shown to be a \sl generalized function \rm
\ci{GS64} of $\3x$ for any fixed $\3y$, meaning that given
any sufficienlty smooth ``test'' function $f(\3x)$, the
integral 
\begin{align}\lab{gen/func}
\la \2\d, f\ra=\int_\rr n\2\d(\3x+i\3y)\,f(\3x)\,d\3x
\end{align} 
defines a bounded linear functional of $f$.  (Generalized
functions are more commonly known as \sl distributions \rm
 \ci{Z65}. We use the former term here in order
to avoid confusion with the term ``source distribution.'')

 By Proposition \ref{T: harmonic}, $\2\d(\3x+i\3y)$
is supported on $\3x\in\5B $ for even $n\ge 4$, and on
$\3x\in  E\60 $ otherwise.  In any case, it has compact
support in the variable $\3x$, and as $\3y\to\30$, this
support contracts to the origin. We will show that
\begin{align*}
\3y\to\30\imp \la \2\d, f\ra\to f(\30), \hb{hence}
 \2\d(\3x+i\3y)\to \d(\3x).
\end{align*} 
To compute the generalized function $\2\d(\3z)$, we first
define the \sl regularized  potential:  \rm 
\begin{align}\label{phi/eps}
\f_\e (\3z)=\q(p-\e )\,\f(\3z), \qq \e >0, \hb{where}
\q\0\x=
\begin{cases}
1 &\text{ if $\x>0$} \\ 
0 & \text{ if  $\x<0$} 
\end{cases}
\end{align}
is the (Heaviside) unit step function.   The regularization
\eq{phi/eps} eliminates the singularities on $  E\60$
and $\5B  $,  replacing them by a discontinuity  in
$\f_\e $ across the spheroid $ E_\e $.   Thus, while
the source of the \sl singular potential \rm $\f(\3z)$ is
concentrated on $E\60  $ or $\5B$,  that of the regularized
potential $\f_\e (\3z)$ is concentrated on $ E_\e    $.  The
advantage gained is that while $\f(\3x+i\3y)$ is infinite for
$\3x\in\5B  \subset  E\60  $, $\f_\e (\3x+i\3y)$
remains bounded in a neighborhood of $ E_\e  $.

Now $\f_\e (\3z)$ vanishes in the interior of $
E_\e  $ but is identical to $\f(\3z)$ in the exterior.
Therefore its source distribution,  defined as a generalized
function by
\begin{align}\label{delta/eps}  
\2\d_\e (\3z) \=\D\f_\e (\3z),
\end{align}
represents an \sl equivalent source distribution \rm
on $\3x\in  E_\e  $  whose potential field simulates that
of $\2\d(\3z)$ in the exterior of $E_\e  $ but vanishes in
the interior.   We call  $\2\d_\e (\3z)$ the \sl regularized
source distribution, \rm and will define the \sl singular \rm
source distribution $\2\d(\3z)$ as the limit of $\2\d_\e (\3z)$
in the sense of generalized functions, for any fixed 
$\3y\in\rr n$. That is,
\begin{align}\lab{delta/limit}
\la \2\d, f\ra\=\lim_{\e \to 0\9+}\la \2\d_\e \,,f\ra
\end{align}
for every test function $f(\3x)$ in $\rr n$. Using 
$\q'\0\x=\d\0\x$, \eq{phi/eps} gives
\begin{align*}
\nabla\f_\e 
=\d(p-\e )\,\f\,\nabla p+\q(p-\e )\nabla\f,
\end{align*}
hence
\begin{align}\label{D/phi/eps}
\D\f_\e =\d'(p-\e )\,\f\,(\nabla p)^2
+2\d(p-\e )\nabla\f\cdot\nabla p
+\d(p-\e )\f\,\D p  +\q(p-\e )\D\f.
\end{align}
 By Proposition \ref{T: harmonic}, $\D\f$ vanishes for 
$p\ge \e /2$, so the last term in \eq{D/phi/eps} vanishes
identically.   The remaining terms show that
$\2\d_\e (\3x+i\3y)$ is indeed  supported on
$ E_\e $ as expected.  Inserting
\begin{align*}
\nabla\f=\f'(\g )\nabla\g=\frac{\g ^{1-n}\,\nabla\g}{\o_n}
\end{align*}
into \eq{D/phi/eps} and using  \eq{orth}--\eq{Du/Dv}, we have
\begin{align}\lab{delta/eps2} 
\2\d_\e (\3z)=\lb \d'(p-\e )\,\f+2\d(p-\e )\,\f'\rb 
\frac{a^2+p^2}{\7\g \,\g }
+\d(p-\e )\,\f\,\frac{n-1}{\7\g \,\g }\,p.
\end{align}
 This expression  represents a generalized function of $\3x$.
To make sense of it we must apply it to a test function
$f(\3x)$ in $\rr n$, assumed to be sufficiently smooth.
(That is,  $f$ is assumed to possess all derivatives
which the ensuing computation requires it to possess. As will
be seen, the required degree of smoothness increases with
$n$.)  For a given value of $\3y\in\rr n$, $\2\d_\e $  acts
on $f$ as in \eq{gen/func},
\begin{align}\lab{action/eps}
\la\2\d_\e \,, f\ra
&=\int_\rr n \2\d_\e (\3x+i\3y)\,f(\3x)\,d\3x.
\end{align}
Using the oblate spheroidal  coordinates 
\eq{OS}, let us write
\begin{align*} 
f(\3x)=f(\r\3\s+\z\,\bh y)=f\v(p,q, \3\s)\=f\v(\g ,\3\s),
\end{align*}
where the two expressions on the right are obtained  by
substituting \eq{p,q} and \eq{rho} for $\z$ and $\r$ in terms
of $(p,q)$.  Let
\begin{align}\label{means}
\7f\v\0\g =\int_{S^{n-2}} f\v(\g , \3\s) \,d\3\s
=\int_{S^{n-2}}f(\r\3\s+\z\,\bh y)\,d\3\s\=\7f(\r,\z). 
\end{align}
The notations $f\v(\g , \3\s)$ and $\7f\v\0\g\=\7f\v(p,q)$ 
are used for convenience and are \sl not \rm meant to imply
analyticity in $\g$.  Because of the normalization
\eq{dsigma}, $\7f\v\0\g$ and  $\7f(\r,\z)$ are the \sl means
\rm of $f\v(\g , \3\s)$ and  $f(\r\3\s+\z\,\bh y)$ over the
sphere $S_\g=E_p\cap H_q=\LB\3x:\  |\3x-\z\3\s|=\r\RB$.
Using the expression \eq{vol/os} for $d\3x$, \eq{action/eps}
becomes
\begin{align*}
\la\2\d_\e \,, f\ra &=
\frac{\o_{n-1}}a
\int_0^\8 dp\int_{-a}^a dq\,\r^{n-3}\,\7\g\,\g\,
\2\d_\e \0\g  \,\7f\v\0\g ,
\end{align*}
 where we have used the fact that $\2\d_\e (\3z)$ in  
\eq{delta/eps2} is independent of $\3\s$ to write it as
$\2\d_\e (p,q)\=\2\d_\e \0\g $.  By \eq{rho},
\begin{align*}
(a^2+p^2)\pl_p\r^{n-3}=(n-3) p\,\r^{n-3}.
\end{align*}
 Hence the first term in   \eq{delta/eps2} gives,
upon  integrating by parts over $p$,
\begin{align*}
&\int_0^\8 dp\int_{-a}^a dq\, 
(a^2+p^2)\r^{n-3}\d'(p-\e )\,\f\0\g \,\7f\v\0\g\\
&\qqq\qqq=-\int_{-a}^a dq\, \r^{n-3}
\lb (n-3)\,\e \,\f\,\7f\v +2\e \f\,\7f\v
+ (a^2+p^2)(\f'\,\7f\v+\f\,\7f\v_p)\rb\\
&\qqq\qqq=-\int_{-a}^a dq\, \r^{n-3}
\lb (n-1)\,\e \,\f\,\7f\v
+(a^2+\e ^2)(\f'\,\7f\v+\f\,\7f\v_p)\rb, 
\end{align*}
where $\7f\v_p=\pl_p\7f\v(p,q)$
and  the integrand is to be evaluated at $p=\e $. Inserting the
other terms in  \eq{delta/eps2} and simplifying, we obtain
\begin{align}\lab{action/eps/2}
\la\2\d_\e \,, f\ra
&=\frac{a^2+\e ^2}a\,\,\o_{n-1}\int_{-a}^a dq\, \r^{n-3}
\lb\f'\,\7f\v-\f\,\7f\v_p\rb.
\end{align}

\begin{prop}\lab{thm:n/ge3} For all $n\ge 3$, the
regularized source distribution $\2\d_\e (\3x+i\3y)$ is 
supported on the oblate spheroid $\3x\in E_\e$, and its
action on a test function $f(\3x)=f\v(\g,\3\s)$ is given by
\begin{align}\lab{action2/eps}
\la\2\d_\e \,,f\ra = I_\e (\7f\v), 
\end{align} 
where $I_\e $ is the linear functional defined by
\begin{align} \lab{I/eps}
I_\e (\7f\v)
=\frac{(a^2+\e ^2)^{\n+1}}{a^{n-2}A_n}\inta
\frac{F\v(\e +iq)}{(\e +iq)^{n-1}}\,dq , \qq\n=\frac{n-3}2\,,
\end{align} 
with 
\begin{gather}\lab{F(gamma)}  
A_n=\frac{\o_n}{\o_{n-1}},\qqq
F\v\0\g =(a^2-q^2)^\n \lb \7f\v\0\g
+\frac{\g\7f\v_p\0\g}{n-2} 
\rb .
\end{gather}
\end{prop}

\sl Proof: \rm This follows immediately from
\eq{action/eps/2}, using 
\begin{align*}
\r^2=\frac{(a^2+\e^2)(a^2-q^2)}{a^2}\, .  \sh2\qed 
\end{align*}

Let us verify that as $\3y\to\30$ and the source
disk shrinks to a point, the source of the singular potential
$\f$ contracts to the usual point source. Letting $q=a\x$ in 
\eq{F(gamma)}, we have 
\begin{align*}
F\v\0\g=a^{n-3}(1-\x^2)^\n
\lb \7f\v\0\g +\frac{\g\7f\v_p\0\g}{n-2}  \rb, 
\hb{where}\g =\e+ia\x.
\end{align*} 
Therefore
\begin{align*}
I_\e (\7f\v)=\frac{(a^2+\e ^2)^{\frac{n-1}2}}{A_n}\int_{-1}^1 
\frac{(1-\x^2)^\n}{(\e +ia\x)^{n-1}}
\lb\7f\v\0\g +\frac{\g\7f\v_p\0\g}{n-2}\rb d\x
\end{align*}
 and
\begin{align*}
\lim_{a\to 0}I_\e (\7f\v)
=\lb\7f\v\0\e+\frac{\e\7f\v_p\0\e}{n-2}\rb
\frac{K_n}{A_n}\,,
\end{align*}
 where
\begin{align*} 
K_n=\int_{-1}^1(1-\x^2)^\n \,d\x
=B\lp\tfrac12, \tfrac{n-1}2\rp
=\frac{\sr{\p}\ \G\lp\frac{n-1}2\rp}{\G\lp\frac n2\rp}
=\frac{\o_n}{\o_{n-1}}=A_n\,.
\end{align*}
 Thus by \eq{action2/eps},
\begin{align*} 
\lim_{a\to 0}\,\la\2\d_\e ,f\ra=\7f\v(\e )
+\frac{\e \7f\v_p\0\e}{n-2}\,.
\end{align*}
 Now let $\e \to 0$, and note that since $E_\e $ contracts to
the origin,
\begin{align*} 
\lim_{\e \to 0}\7f\v\0\e =
\lim_{\e \to 0}\int_{S^{n-2}}f\v(\e ,\3\s) d\3\s=f(\30),
\hb{hence} \lim_{\e \to 0}\,\lim_{a\to 0}\,\la\2\d_\e ,f\ra
=f(\30).
\end{align*}
If we assume that the two limits can be exchanged (as will
be verified later) and use the definition \eq{delta/limit} of
$\2\d$, this states that $\lim_{a\to 0}\,\la\2\d,f\ra=f(\30)$,
giving the following important result.

\begin{thm}\lab{T:point/source} The singular source
distribution $\2\d(\3z)$ is an  extension of the
usual point-source distribution in $\rr n$, in the sense that
\rm \begin{align}\lab{pt/source}
\3y\to \30\imp \2\d(\3x+i\3y)\to\d(\3x). \sh2\qed
\end{align}

\end{thm}

\section{Singular Source Distributions}
\lab{S: SingSource} 

We are ready at last compute the  singular source
distributions.  By \eq{delta/limit} and 
\eq{action2/eps},  they  given by  the limit
\begin{align}\lab{/2/deta}
\la\2\d, f\ra=\lim_{\e \to0}I_\e (\7f\v).
\end{align}
However,  we cannot simply let $\e =0$ in the expression 
\eq{I/eps} for $I_\e (\7f\v)$, since  the resulting integral
diverges.  Recall the decomposition of $E\60$ given in
Equation \eq{decomp/E0}: 
\begin{align}\lab{decomp2/E0}
E\60=  E\60\9+\cup  E\60\9-\cup\5B\,,
\end{align} 
where $ E\60\9\pm$ are the interiors of the front 
and back sides of $ E\60$ and $\5B$ is its rim,  the branching
sphere. This decomposition recognizes the nature of $ E\60$
as a limit of closed ellipsoids. Although the two open disks
$ E\60\9\pm$ look identical to a \sl continuous \rm function,
they look distinct to a generalized function like $\2\d$
which is singular across $ E\60$. Furthermore, the
oblate spheroidal  coordinates are an ideal tool for resolving
this decomposition  since $q>0$ on $ E\60\9+$, $q<0$ on 
$E\60\9-$ and $q=0$ on $\5B$. We will compute $\la\2\d,
f\ra$ by decomposing  the integral $I_\e(\7f\v)$ in a way
similar to \eq{decomp2/E0} and then taking the limit
$\e\to0$.  The integral over $ E\60\9+\cup  E\60\9-$  gives a
sum of \sl single and double layer distibutions \rm of
dimension $n-1$ over the interior of the disk $ E\60$, while
the integral over $\5B$ gives a \sl boundary  distribution \rm
of dimension $n-2$  over $\5B$.  All these distributions are
well-defined, giving a finite expression for $\la\2\d, f\ra$. 
  
\sl The divergence of $I_0(\7f\v)$ is therefore
caused by  the attempt  to represent the boundary
distribution on $\5B$ as part of an integral of dimension
$n-1$ over $E\60$, and the above regularization simply
amounts to recognizing this fact. \rm  

To regularize the integral
\begin{align} \lab{I/eps/2}
I_\e (\7f\v) =\frac{(a^2+\e ^2)^{\n+1}}{a^{n-2}A_n}
\inta\frac{F\v(\e +iq)}{(\e +iq)^{n-1}}\,dq ,
\end{align} 
define the Taylor coefficients
\begin{align}\lab{Tm}
T_m\0\e=\frac1{m!}\pl_q^m F\v(\e+iq)\bigm|_{q=0}\,, \qqq
 T_m\=T_m\00
\end{align} 
and the Taylor polynomials approximating $F\v(\e +iq)$ and
$F\v(iq)$ to order $q^{n-2}$,
\begin{align}\lab{F_(n-2)}
F\v_{n-2}(\e, q)=\sum_{m=0}^{n-2}q^m T_m\0\e,\qq
F\v_{n-2}(q)=\sum_{m=0}^{n-2}q^m T_m\,.
\end{align} 
We rewrite the integral in
\eq{I/eps/2} as
\begin{align}\lab{reg}
\inta \frac{F\v(\e+iq)}{(\e+iq)^{n-1}}\, dq
=\inta \frac{F\v(\e+iq)-F\v_{n-2}(\e, q)}{(\e+iq)^{n-1}}\, dq
+\sum_{m=0}^{n-2}i^{-m}T_m\0\e \l^m_k\0\e, 
\end{align} 
where
\begin{align} \lab{lam/km}
\l^m_k\0\e&=
\inta\frac{i^mq^m dq}{(\e+iq)^k}, \qq 0\le m< k.
\end{align}
If  $f(\3x)$ is  differentiable to order $n-2$,
so is $F\v(\e+iq)\=F\v(\e, q)$. Then
\begin{align*}
F\v(\e+iq)-F\v_{n-2}(\e, q)=O(q^{n-1})
\end{align*} 
and the integral on the right-hand side of \eq{reg} has a
finite limit as $\e\to0$. It therefore  remains to compute the
limit of the sum. We begin by finding
\begin{align} \lab{lam/10}
\l\90_1\0\e&=\inta\frac{dq}{\e+iq}
=2\e\int_0^a \frac{dq}{\e^2+q^2}
=2\tan\inv\frac a\e
=\p-2\tan\inv\frac\e a\=\a\0\e.
\end{align}
All the other integrals  \eq{lam/km} can be computed from
the recursion relations
\begin{align*}
\l\90_{k+1}\0\e&=-\frac1k\pl_\e  \l\90_k\0\e
=\frac{(-1)^k}{k!}\pl_\e ^k\a\0\e\\
\l^{m+1}_k\0\e
&=i^m\inta\frac{(\e+iq-\e)q^mdq}{(\e+iq)^k}
=\l^m_{k-1}\0\e-\e \l^m_k\0\e.
\end{align*}
 We are interested in the limit $\e\to 0$, where
these relations imply
\begin{gather}\lab{lam/k/2}
\l_{k+1}\=\l_{k+1}\90\00
=\frac{(-1)^k}{k!}\pl_\e ^k\a\00\=(-1)^k L _k\\
\l^m_k\00=\l\90_{k-m}\00=\l_{k-m}\,, \qq 0\le m< k.
\lab{lam/km/rec}
\end{gather}
The Taylor coefficients $L_k$ of $\a\0\e$ are
obtained from the expansion
\begin{align}\lab{taylor}
\a\0\e=\p-2\tan\inv\frac\e a
=\p+2\sum_{l=1}^\8\frac{(-1)^l}{2l-1}
\frac{\e^{2l-1}}{a^{2l-1}}.
\end{align}
 Thus
\begin{align}\lab{lam/k/final}
\l_1=\p, \qq
\l_{2l}=\frac{2(-1)^{l+1}}{(2l-1)a^{2l-1}}, \qq 
\l_{2l+1}=0,\qq l\ge 1.
\end{align}
 By \eq{lam/km/rec}, this gives finite values for all the
coefficients $\l_k^m\00$.  Note that the original
expression \eq{lam/km} diverges if we set $\e=0$. The
finiteness of the limits depends on delicate cancellations of
contributions from positive and negative values of $q$ when
$\e>0$, just as happens in Cauchy's principal
value integral. In fact, \eq{lam/10} shows that \sl the present
regularization  reduces to the principal value integral when
$k=1$. \rm

Using \eq{reg}, we can now compute the limit   $\e\to 0$ in
\eq{/2/deta}:
\begin{align}\lab{deta/2} 
\la \2\d, f\ra
&=V_n\0f+\frac a{A_n}
\sum_{m=0}^{n-2}i^{-m}\,T_m\,\l_{n-m-1}\,,\\
V_n\0f&\= \frac{i^{1-n}a}{A_n}
\inta \frac{ F\v(iq)-F\v_{n-2}(q)}{q^{n-1}}\,dq . \lab{Vn}
\end{align} 
$V_n\0f$ will be shown to be a \sl bounded linear functional
\rm of $f$, and this establishes $\2\d$ as a well-defined
generalized function.

Since $\7f\v\00$ is the mean of $f$ over
$\5B$,  the terms in the sum in \eq{deta/2}  represent means
of $f$ and its derivatives over $\5B$. On the other hand,
\eq{Vn} represents an integral of $f$ and its normal
derivative over the \sl interior \rm $ E\60\9+\cup 
E\60\9-$, since the boundary terms have already been
subtracted in the form of $F\v_{n-2}\0q$. Thus,  \eq{deta/2}
is the promised decomposition of the source distribution
corresponding to  \eq{decomp2/E0}.

Equation \eq{deta/2}  can be greatly simplified because of
certain symmetries satisfied by $\7f\v(iq)$. We claim that
\begin{align}\label{sym}
\7f\v(-iq)=\7f\v(iq), \qq
\7f\v_p(-iq)=-\7f\v_p(iq), \qq
\7f\v_q(-iq)=-\7f\v_q(iq). 
\end{align}
To see this, note that the coordinates 
$\g =\pm iq$ denote \sl the same point \rm of $ E\60$,
regarded as belonging to $ E\60\9\pm$. Since the test
function $f(\3x)$ is continuous across $ E\60$, so is its
integral $\7f\v(iq)$, and this proves the first relation.
On the other hand,  $p$  increases in the $\pm \z$ direction on
$ E\60\9\pm$, which proves the second relation.  Finally,
the third relation follows from the first
by differentiation.

By  \eq{sym}, the function
\begin{align}\lab{F(iq)} 
F\v(iq)=(a^2-q^2)^\n\lb\7f\v(iq)
+i\,\frac{q\7f\v_p(iq)}{n-2}\rb 
\end{align}
 is \sl even, \rm so its odd Taylor coefficients vanish:
\begin{align}\lab{Tm=0}
m\text{ is odd }&\imp T_m=0.
\end{align}
 Furthermore,  \eq{Vn} shows that
\begin{align}\lab{Vn=0}
n\text{ is even }&\imp V_n\0f=0.
\end{align}

Considering the cases of odd and even $n$ separately and
inserting the values of $\l_k$ from   \eq{lam/k/final}
gives the following result.

\begin{thm} \lab{T:source} 
The singular source distribution $\2\d(\3z)$ is a bounded
linear functional which acts on a test function $f\in
C^{n-2}(\rr n)$ as follows.  For even
$n=2k+2\ge 4$, 
\begin{align}\lab{even}
\la\2\d, f\ra=\frac{\p a}{A_n}\,(-1)^k T_{2k}\,,
\end{align}
where $A_n$ and $T_{2k}$ are given by \eq{F(gamma)}  and 
\eq{Tm}. For odd $n=2k+3\ge 3$,
\begin{align}\lab{odd}
\la \2\d, f\ra =
V_n\0f+\frac{2(-1)^k}{A_n}\sum_{l=0}^k
\frac{a^{2l-2k}\,T_{2l}}{2k-2l+1}\,,
\end{align}
where $V_n\0f$ is given by \eq{Vn}.

\end{thm}

\sl Proof: \rm
To prove \eq{even}, note first
that  $V_n\0f$   vanishes by \eq{Vn=0}. Furthermore,
the sum in  \eq{deta/2} reduces to a single term
because $T_m\l_{n-m-1}=0$ unless $m$ is even,
which implies that $n-m-1$ is odd; but the  only 
nonvanishing coefficient $\l_l$ with odd $l$ is $\l_1=\p$.

Equation \eq{odd} follows directly from \eq{deta/2},
\eq{lam/k/final} and \eq{Tm=0}. To establish $\2\d$ as a
distribution,  we must still prove that $V_n\0f$ is a
bounded linear functional for odd $n$. Since the integrand in
\eq{Vn} is even when $n$ is odd, we have
\begin{align*}
V_n\0f= \frac{2i^{1-n}a}{A_n}\int_0^a
 \frac{F\v(iq)-F\v_{n-2}(q)}{q^{n-1}}\,dq .
\end{align*} 
By Taylor's theorem,
\begin{align*}
F\v(iq)-F\v_{n-2}(q)=\frac1{(n-2)!}\int_0^q(q-w)^{n-2}
(F\v)^{(n-1)}(iw)\,dw.
\end{align*} 
Therefore
\begin{align*}
|F\v(iq)-F\v_{n-2}(q)|\le q^{n-1}M, \qqq
M=\frac{\max |(F\v)^{(n-1)}(iw)|}{(n-2)!},
\end{align*} 
and
\begin{align*}
|V_n\0f |\le 2a^2M.
\end{align*} 
Since $M$ depends boundedly on $f$, this shows that
$V_n\0f$  is a bounded linear functional as claimed. \sh2\qed

Thus, for even $n\ge 4$,
the integral over the interior of  $ E\60$ vanishes
and the singular source is concentrated on  $\5B$. This is to
be expected, since  $\g ^{2-n}$  has no  branch cut  and
therefore the only singularity occurs on the boundary. 

We can now derive a useful expression for the source
distribution $\2\d(\3x+i\3y)$ in the cylindrical coordinates
$(\r,\z,\3\s)$ adapted to $\3y\ne\30$. As a byproduct, it
will be seen that for all $n\ge 3$, the test function
$f(\3x)$ need only be in $C^k(\rr n)$ with $k=[\frac{n-1}2]$,
instead of $C^{n-2}(\rr n)$ as assumed in Theorem
\ref{T:source}.

Our first task is to
rewrite the function $F\v(iq)$ in \eq{F(iq)}  in terms of
cylindrical coordinates. Recall that
\begin{gather}
\z =\frac{pq}a, \qqq
\r =\frac{{\sr{(a^2+p^2)(a^2-q^2)}}}a \lab{cylCS}\\
\3x=\r\3\s+\z\,\bh y, \qq\3\s\in S^{n-2}\perp\3y,\notag\\
f(\r\3\s+\z\,\bh y)=f\v(p,q,\3\s). \lab{def/h}
\end{gather}
Then the mean of $f$ on $S_\g=E_p\cap H_q$ becomes
\begin{align}\label{cyl/mean}
\7f\v(p,q)
=\int_{S^{n-2}} h(\r\3\s+\z\,\bh y)\,d\3\s\=\7f(\r,\z),
\end{align}
which is  the mean of $f$ on the sphere 
$S_{\r,\z}=\{\3x:  |\3x-\z\,\bh y|=\r\}$.

It will be shown elsewhere \ci{K00} that the source
distributions for odd $n$ can be derived from those for even
$n$ by integrating over one of the coordinates. Specifically,
we have:

\begin{thm}\label{T:descent}
Let $n\ge 3$, and denote points in $\4C^{n+1}$ by
\begin{align*}
\/z=\/x+i\/y=(\3z, s+it), \qqq \3z\in\cc n.
\end{align*}
For fixed $\/y\ne\30$, the singular source distributions
$\2\d_n(\3x+i\3y)$ in $\rr n$ and
$\2\d_{n+1}(\/x+i\/y)$ in $\4R^{n+1}$ are related by
\begin{align}\label{descent}
\2\d_n(\3z)=\i1s \2\d_{n+1}(\3z, s).
\end{align}
That is, for a test function $f(\3x)$ in $\rr n$ we have
\begin{align}\label{descent2}
\la \2\d_n\,, f\ra=\i1s \int_\rr n d\3x\ 
\2\d_{n+1}(\3x+i\3y, s) f(\3x).
\end{align}

\end{thm}

For $\3y=\30$, \eq{descent} follows from 
$\d_{n+1}(\3x, s)=\d_n(\3x)\,\d\0s$. For $\3y\ne \30$, this
tensor product decomposition fails but Theorem
\ref{T:descent} still holds. We therefore state the following
theorem for the simpler case of
\sl even \rm $n\ge 4$, though the transformation to
cylindrical coordinates derived below will be used later in
the explicit computation of the extended source in $\rr3$.

\begin{thm} \lab{T:cylin}
For $n=2k+2\ge 4$,  the source distribution $\2\d$ is a
bounded linear functional on test fucntions 
 $f(\3x)$ in $C^k(\rr n)$ whose action is given in
the cylindrical coordinates $(\r,\z,\3\s)$ by
\begin{align}\label{cyl}
\la\2\d, f\ra
=\frac {a\sr{\p}}{\G(k+\tfrac12)}\,D_\r^k
\,F\0\r\Bigm|_{\r=a}\,,
\end{align}
\end{thm}
where
\begin{gather}
D_\r=\ptl{}{(\r^2)}=\frac1{2\r}\ptl{}\r\,,\qq
F\0\r=\r^{2k-1}\lb
\7f(\r,0)+i\,\frac{a^2-\r^2}{2ka}\,\7f_\z(\r,0)\rb.
\lab{H}
\end{gather}

\sl Proof: \rm 
By \eq{cylCS}, we have
\begin{align}\lab{fp}
\7f\v_p(p,q)=\frac{\r p}{a^2+p^2}\,\7f_\r(\r,\z)
+\frac qa\,\7f_\z(\r,\z).
\end{align}
Since
\begin{align*}
p=0\imp q^2=a^2-\r^2,
\end{align*}
it follows from \eq{F(iq)} that 
\begin{align}\lab{H(rpho)}
F(iq)&=(a^2-q^2)^\n\lb\7f\v(0,q)
+i\,\frac{q\7f\v_p(0,q)}{n-2}\rb\notag\\
&=\r^{n-3}\lb
\7f(\r,0)+i\,\frac{q^2}{(n-2)a}\,\7f_\z(\r,0)\rb=F(\r).
\end{align}
Note that \eq{H(rpho)} makes sense only because $F(iq)$ is \sl
even \rm and so does not depend on the sign of
$q$, which cannot be recovered from $\r$ when $\z=0$ since
$\sgn q=\sgn \z$.
With
\begin{align*}
D_q=\ptl{}{(q^2)}=\frac1{2q}\ptl{}q=-D_\r 
\qq\text{when}\qq p=0,
\end{align*}
we have
\begin{align*}
F(iq)=\sum_{m=0}^\8 T_{2m}\,q^{2m} \imp
T_{2m}=\frac1{m!}\,D_q^m \,F(iq)\Bigm|_{q=0} 
=\frac{(-1)^m}{m!}\,D_\r^m\, F(\r)\Bigm|_{\r=a}\,.
\end{align*}
For $n=2k+2\ge 4$,  
\begin{align*}
\o_n=\frac{2\p^{k+1}}{\G(k+1)} =\frac{2\p^{k+1}}{k!},\qq
\o_{n-1}=\frac{2\p^{k+\tfrac12}}{\G(k+\tfrac12)} ,
\end{align*}
and \eq{even} gives
\begin{align*}
\la\2\d, f\ra=\frac{\p a\o_{n-1}}{\o_n}\,(-1)^kT_{2k}
=\frac {a\sr{\p}}{\G(k+\tfrac12)} \,D_\r^k
\,F\0\r\Bigm|_{\r=a}\,.
\end{align*}
Now $F\0\r$ already contains one derivative ($\7f_\z$)
and $D_\r^k$ computes $k$ more,  hence it suffices to have 
$f\in C^{k+1}(\rr n)$. But the highest derivative of $f$
occuring in \eq{cyl} is
\begin{align*}
\frac{\r^{2k-1}}{(2\r)^k}\,\cdot
i\frac{a^2-\r^2}{2ka}\,\pl_\r^k\7f_\z(\r,0),
\end{align*}
which vanishes at $\r=a$. Hence the highest nonvanishing
derivative is of order $k$, so it suffices for $\7f(\r,\z)$ to be
$k$ times continuously differentiable as claimed, and so does
$f(\3x)$. \sh2\qed 

An application of  \eq{descent2} now shows that for 
$n=2k+1$, we need $f\in C^k(\rr n)$ in order for
$\la\2\d_n, f\ra$ to make sense. Thus for \sl any \rm
$n\ge 3$, $f$ needs to be in
$C^k(\rr n)$ with 
\begin{align*}
k=\lb\frac{n-1}2\rb=
\begin{cases}
\frac{n-2}2, & n\text{ even}\\
\frac{n-1}2, & n\text{ odd}.\\
\end{cases}
\end{align*}

\section{Computation of Sources in $\rr3$ and $\rr4$}
\lab{S: CompInR2R3}

We now compute the singular source distribution in $\rr3$
explicitly and interpret the result. Equation \eq{F(iq)}
becomes
\begin{align}\lab{F/n=3}
F\v(iq)=\7f\v(iq)+iq\7f\v_p(iq),
\end{align}
hence $T_0=\7f\v\00$, $T_1=0$, and \eq{F_(n-2)} becomes
\begin{align*} 
F\v_1(q)=\7f\v\00.
\end{align*} 
Equation \eq{odd} therefore gives
\begin{align*}
\la \2\d, f\ra=
- a\int_{-a}^a\frac{\7f\v(iq)
+iq\7f\v_p(iq)-\7f\v\00}{q^2}\,dq +\7f\v\00.
\end{align*}
Using the symmetry of the integrand, we obtain the
following  result.

 \begin{prop}  \lab{T: source/n=3}
The singular source  distribution
\begin{align*}
\2\d(\3x+i\3y)=-\D\frac1{4\p\g }, \qq \3x+i\3y\in\cc3,
\end{align*} 
is a bounded linear functional whose action on a test function
$f(\3x)=f\v(\g,\3\s)$ in $C^1(\rr3)$, is given 
in oblate spheroidal and cylindrical coordinates by
\begin{align}\lab{singular/source/n=3}
\la\2\d, h\ra=L_0+L_1+iL_2\,,
\end{align}
where
\begin{align*}
L_0&=\7f\v\00=\7f(a,0)\\
L_1&=-a\int_0^a\frac{\7f\v(iq)-\7f\v\00}{q^2}\,dq
=-a\int_0^a
\frac{\7f(\r,0)-\7f(a,0)}{(a^2-\r^2)^{3/2}}\,\r\,d\r\\
L_2&=-a\int_0^a\frac{\7f\v_p(iq)}q\,dq
=-\int_0^a\frac{\7f_\z(\r,0)}{\sr{a^2-\r^2}}\,\r\,d\r. 
\end{align*}

\end{prop}

\sl Proof: \rm The action in oblate spheroidal
coordinates follows immediately from \eq{odd}. To obtain the
action in cylindrical coordinates, use \eq{fp}. \sh2\qed
  
Note that since
\begin{gather*}
\lim_{a\to 0}\7f(a,0)=  f(\30) \hb{and}
\lim_{a\to 0}L_1=\lim_{a\to 0}L_2=0,
\end{gather*} 
\eq{singular/source/n=3} shows that
\begin{align*}
\3y\to\30\imp\2\d(\3x+i\3y)\to\d(\3x).
\end{align*} 
This was already seen in Theorem \ref{T:point/source} for
all $n\ge 3$, but that proof was less rigorous because it
depended on the assumption that the order of the limits $a\to
0$ and $\e\to 0$ can be exchanged.

We now state some other interesting properties
of the expression \eq{singular/source/n=3} which will help
interpret its three terms.

\begin{prop}
\label{T: charge/dipole/n=3} The monopole and dipole
moments of $\2\d(\3x+i\3y)$ in
$\rr3$ are
\begin{align*}
Q\=\int_\rr3\2\d(\3x+i\3y)\,d\3x=1,\qqq
\3P\=\int_\rr3\3x\,\2\d(\3x+i\3y)\,d\3x
=-i\3y.
\end{align*}
Given a point source with \rm general \it complex coordinates
$\3z\6S=\3x\6S+i\3y\6S\in\cc 3$,
the \rm centroid \it of its charge distribution is
\begin{align}\label{centroid}
\3C(\3z\6S)\=\int_\rr3\3x\,\2\d(\3x-\3z\6S)\,d\3x
=\3z\6S\,. 
\end{align}
\end{prop}

\sl Proof: \rm  To compute $Q$, apply Proposition 
\ref{T: source/n=3} with $f(\3x)\=1$. To  find $\3P$, apply it
to the \sl vector-valued \rm test function
\begin{align*}
\3f(\3x)=\3x=\r\3\s+\z\bh y.
\end{align*} 
Finally,
\begin{align*}
\3C(\3z\6S)
=\3x\6S\int_\rr3\,\2\d(\3x-\3x\6S-i\3y\6S)\, d\3x
+\int_\rr3(\3x-\3x\6S)\,
\2\d(\3x-\3x\6S-i\3y\6S)\, d\3x
=\3x\6S+i\3y\6S=\3z\6S\,.\sh2\qed
\end{align*}
Equation \eq{centroid} is a natural extension to
$\cc3$ of the formula
\begin{align*}
\int_\rr3\3x\,\d(\3x-\3x\6S)\, d\3x=\3x\6S,\qqq
\3x\6S\in\rr3. 
\end{align*} 
Proposition \ref{T: charge/dipole/n=3}  sheds some
light on  the nature of the source distribution $\2\d$. 
\begin{itemize} 
\item 
$L_0$ is the mean of $f$ over the rim $\5B$. Since
$L_1=L_2=0$ when $f(\3x)\=1$, we see that the ``charge'' $Q$
resides entirely on this rim. \rm
\item 
$L_1$ is an integral of $f$ over
$E\60$ which does not involve its derivatives, thus
representing a \sl single layer distribution \rm on $E\60\,$.
Actually, since the contributions from the rim are
subtracted, the single layer resides on the \sl interior \rm 
of $E\60\,$.
\item
$L_2$ is an integral of the normal derivative of
$f$ over $E\60\,$, so it represents a \sl
double layer distribution \rm  which may be regarded as
equal and opposite charge distributions on $E\60\9+$ and 
$E\60\9-$.  This  is confirmed by $\3P=-i\3y$.

\end{itemize}

\begin{prop}   The source distribution in $\cc 4$,
\begin{align*}
\2\d(\3z)=-\D\frac1{4\p^2\g ^2}\,, 
\end{align*} 
acts on a test function $f\in C^1(\rr4)$ in cylindrical
coordinates as follows:
\begin{align}\lab{delta/n=4/cyl}
\la\2\d, f\ra=\7f(a,0)+a\7f_\r(a,0)-ia\7f_\z(a,0). 
\end{align}
\end{prop}

\sl Proof: \rm From \eq{H} we find
\begin{align*}
F\0\r
&=\r\lb \7f(\r,0)+i\,\frac{a^2-\r^2}{2a}\,\7f_\z(\r,0)\rb,
\end{align*}
and \eq{cyl} becomes $\la\2\d, f\ra=\pl_\r F\0a$, which
gives \eq{delta/n=4/cyl}. \sh2\qed 

As $a\to 0$, this gives $\la\2\d, f\ra\to f(\30)$, which
confirms that $\2\d(\3x+i\3y)\to\d(\3x)$.  This property and 
$Q=1, \3P=-i\3y, \3C(\3z\6S)=\3z\6S\,$, as in  Proposition 
\ref{T: charge/dipole/n=3}, can be shown to hold for \sl all
\rm $n$ \ci{K00}.

The holomorphic potential in $\cc4$ was derived from a
different point of view in \ci[Section 11.2]{K94}, where it
was shown to decompose into \sl causal \rm (retarded) and
\sl anticausal \rm (advanced) ``physical wavelets.''  These are
closely related to the \sl complex-source pulsed beams \rm
in the engineering literature \ci{HF89}.

\section{Connection to Spacetime and Wave Equations}
\lab{S: Waves}

It was proposed in Section 1  that point sources in physics be
replaced by extended sources based on a continuation of
physics to complex spacetime.  The distribution $\2\d(\3z)$ in
complex space ($\cc n$) seems like a promising model, since it
is supported on the disk $E\60$ (when $n$ is odd) or the sphere
$\5B$ (when $n$ is even). But to do physics we need to add \sl time \rm to this
complex-space formalism.  We will first give an argument
suggesting that time is \sl already \rm included in
$\4C^{n+1}$, then justfy this by proving that with this
interpretation, the extended source distribution
$\2\d(\/z)$ in $\4C^{n+1}$ acts as a \sl propagator \rm in
spacetime, generating solutions of the  Cauchy problem for
the wave equation from their initial values.

Let  us write
\begin{align*}
\/z&=\/x+i\/y=(\3z,s+it)\in \4C^{n+1},\kn2 &&\3z\in\cc n\\
\/x&=(\3x, s),\qq \/y=(\3y, t)\in\4R^{n+1},\kn2  && |\3x|=r.
\end{align*}
Analytic potential theory in $\4C^{n+1}$ is based on the
complex distance function $\g(\/z)$ defined by
\begin{align*}
\g^2&\=\/z^2=\3z^2+(s+it)^2,\qqq \re\g\ge 0,
\end{align*}
which reduces to the \sl Euclidean \rm metric in $\4R^\+n1$
when $\/y\to\30$,
\begin{align*}
\g^2\to\/x^2=r^2+s^2,
\end{align*}
and to the \sl Lorentzian \rm metric in  $\4R^{n, 1}$ when 
$s\to 0$ and $\3y\to\30$,
\begin{align*}
\g^2\to r^2-t^2.
\end{align*}
This suggests that  \sl the  imaginary part $t$ of the complex
space coordinate $z_{n+1}=s+it$  should be interpreted
as time. \rm  For this reason,  we refer to its real part $s$
as the \sl Euclidean time. \rm    
 
The idea of time as an imaginary space coordinate dates back
to Minkowski \ci{M23},  who realized in 1908 that
Einstein's new special relativity theory can be based on a
unified four-dimensional spacetime.  More recently, complex
spacetime has become an important concept in quantum field
theory \ci{SW64, GJ87}, twistor theory \ci{PR86} and
string theory \ci{P98}.  Even so, it is generally
regarded as a useful mathematical tool rather than a
fundamental aspect of physical reality.  For example, see the
discussion in
\ci[p.~51]{MTW73}.

In previous work \ci{K77, K78, K87,K90, K94}, I have
attempted to make complex spacetime ``concrete'' by giving 
detailed physical interpretations of the imaginary as well as
the real coordinates. This is also the present motivation for
developing holomorphic potential theory, where the
gemetrical and physical significance of $\3y$ has been
emphasized. (The \sl physical \rm  significance of $\3y$ is
related to the  directivity   of the physical wavelets
associated with $\f(\3x+i\3y)$ \ci{HF89, K94, K00}.)

Assume for simplicity that the Euclidean world is
completely democratic with respect to the real space
variables $\3x\in\rr n$ and the Euclidean time $s$, so that
there is no prefered direction in $\4R^{n+1}$. Then if
$\/y\ne\30$, we may choose a coordinate system in which
\begin{align*}
\/y=(\30, t), \qq t=|\/y|>0, \qqq \/z=(\3x, s+it)\in\4C^\nn.
\end{align*}
In that case, the cylindrical coordinates in $\4R^\+n1$ are
\begin{align*}
\z=\/x\cdot \rh y=s, \qqq
\r=\sr{\/x^2-\z^2}=r,
\end{align*}
and a vector $\/x\in\4R^\+n1$ is represented by
\begin{gather*}
\/x=r\3\s+s\rh y,\qqq \3\s\in S^{n-1}\subset \rr n,
\end{gather*}
where we do not distinguish between $\3\s\in\rr n$ and
$(\3\s,0)\in\4R^\+n1$.  
As before, we denote by $d\3\s$  the
normalized surface measure on $S^{n-1}$, so that $\int
d\3\s=1$ and
\begin{align*}
\7f(r,s)\=\int_{S^{n-1}} f(r\3\s+s\rh y)\,d\3\s
\end{align*}
is the  mean of $f(\3x, s)$ over the sphere $|\3x|=r$. 

For simplicity, we assume to begin with that the dimension 
$n$ of space is \sl odd, \rm so that $n+1$ is even and \eq{cyl}
can be applied to $\2\d(\/z)$.  We will later  extend our
results to \sl even \rm values of $n$ by applying
\eq{descent2}.   Using \eq{cyl} with the substitutions  
\begin{align}\label{subst}
n\to n+1\=2k+2,\qq\r\to r,\qq \z\to s,\qq a\to t
\end{align}
gives the action of $\2\d(\/x+i\/y)$ on a test function
$f(\/x)$:
\begin{align}\label{action/n+1}
\la\2\d, f\ra
=\frac {t\sr{\p}}{\G(k+\tfrac12)}\, D_r^k F\0r\Bigm|_{r=t},
\end{align}
where
\begin{align}\label{Dr,H}
D_r=\frac1{2r}\,\ptl{} r\,,\qq
F\0r=r^{2k-1}\lb \7f(r,0)
+i\,\frac{t^2-r^2}{2k t}\,\7f_s(r,0)\rb.
\end{align}
To establish a connection with the wave equation, define the
function $\2f(\/z)$ on $\4C^\nn$ by the convolution
\begin{align}\label{conv}
\2f(\/z)\=\int_{\4R^\+n1} \2\d(\/z-\/x') \,f(\/x')\,d\/x'.
\end{align}
This is an \sl extension \rm of $f$ from $\4R^{n+1}$ to
$\4C^\nn$, since
\begin{align}\label{ext}
\/y\to\30\imp \2\d(\/x+i\/y-\/x')\to\d(\/x-\/x')\imp
 \2f(\/x+i\/y)\to f(\/x).
\end{align}
Now 
\begin{align*}
\g(-\/z)=\g(\/z)\imp\f(-\/z)=\f(\/z)\imp
\2\d(-\/z)=\2\d(\/z),
\end{align*}
hence
\begin{align}\label{htilde}
\2f(\/x+i\/y)
=\int_{\4R^\+n1} \2\d(\/x'-\/x-i\/y)\,f(\/x')\,d\/x'
=\int_{\4R^\+n1}
\2\d(\/x'-i\/y)\,f^\/x(\/x')\,d\/x'  =\la \2\d, f^\/x\ra,
\end{align}
where $ f^\/x$ is the test function defined by
\begin{align*}
f^\/x(\/x')=f(\/x+\/x')=f(\3x+\3x', s+s').
\end{align*}
Note that in \eq{htilde}, it is $\2\d(\/x'-i\/y)$ rather than
$\2\d(\/x'+i\/y)$ that acts on $f^\/x$. We will 
account for this in Equation \eq{htilde2} below by letting
$i\to -i$ in \eq{Dr,H}. Thus, since $\/z=\/x+i\/y=(\3x, s+it)$,
\eq{action/n+1} gives
\begin{align}\label{htilde2} 
\2f(\3x, s+it)
=\frac {t\sr{\p}}{\G(k+\tfrac12)}
\, D_r^k F^\/x\0r\Bigm|_{r=t}, \qqq
F^\/x\0r=r^{2k-1}\lb \7f^\/x(r,0)
+i\,\frac{r^2-t^2}{2k t}\,\7f^\/x_{s'}(r,0)\rb
\end{align} 
with
\begin{align*}
\7f^\/x(r,s')\=\int_{S^{n-1}}f^\/x(r\3\s+s'\rh y)\,d\3\s
=\int_{S^{n-1}}f(\/x+r\3\s+s'\rh y)\,d\3\s
=\int_{S^{n-1}}f(\3x+r\3\s, s+s')\,d\3\s.
\end{align*}
Therefore
\begin{align}\lab{hx}
\7f^{\/x}(r,0)=\int_{S^{n-1}}f(\3x+r\3\s, s)\,d\3\s
\end{align}
is the mean of $f$ over the sphere of radius $r$ centered
at $\/x$ and orthogonal to $\/y$, and
\begin{align}\lab{hxs}
\7f^{\/x}_{s'}(r,0)\=\pl_{s'}\7f^{\/x}(r,s')\Bigm|_{s'=0}
=\int_{S^{n-1}} f_s(\3x+r\3\s, s)\,d\3\s\=\7f_s^\/x(r,0)
\end{align}
is the mean of $f_s$ over the same sphere.

\begin{thm}\label{T:cauchy} 
Let $n=2k+1\ge 3$ and $f\in C^{k+2}(\4R^{n+1})$. Then
$\2f(\3x, s+it)$ belongs to $C^2(\4R^{n,1})$ as a function of
$(\3x, t)$, and it is the unique ``classical'' solution to the 
following Cauchy problem for the wave equation:
\begin{subequations}\label{ic}
\begin{gather}
\pl_t^2\,\2f(\3x, s+it)=\D_\3x\,\2f(\3x, s+it)\\
\lim_{t\to 0}\2f(\3x, s+it)=f(\3x,s), \qq
\lim_{t\to 0}\pl_t\2f(\3x, s+it)=if_s(\3x, s).
\end{gather}
\end{subequations} 
Furthermore, if $f(\3x, s)$ can be  continued analytically in
$s$, then $\2f(\3x, s+it)$ is that continuation.

\end{thm}

\sl Proof: \rm 
The first of the initial conditions is just
the extension property \eq{ext}, which has already been
established. In view of this, the second initial condition can
be written as
\begin{align}\label{cr}
\lim_{t\to0}\lp\pl_s+i\pl_t\rp\2f(\3x, s+it)=0, 
\end{align} 
which is the \sl Cauchy-Riemann equation \rm at $t=0$.
If \eq{cr} were to hold in a  complex neighborhood 
$U$ of $s_0$,  it would imply that $\2f(\3x, s+it)$ is
holomorphic  in  $s+it\in U$. Since analytic continuation is
unique when it exists, this proves our claim that $\2f(\3x,
s+it)$ is the analytic continuation of $f(\3x, s)$ whenever
the latter exists.

\sl Without \rm assuming analyticity, we now prove that
$\2f(\3x, s+it)$ solves the above Cauchy problem. Let
\begin{gather}\label{u,f,g} u(\3x, t)=\2f(\3x, s+it),\qq
v(\3x, r)=\7f^\/x(r, 0), \qq w(\3x, r)=i\7f^\/x_s(r, 0),
\end{gather}
where the dependence on the parameter $s$ is suppressed. 
Then \eq{htilde2}  becomes
\begin{align}\label{poisson}
u(\3x, t)=\frac {t\sr{\p}}{\G(k+\tfrac12)}\, D_r^k 
\lb  r^{2k-1} v(\3x,r)
+\frac{r^2-t^2}{2kt} \, r^{2k-1}w(\3x,r)\rb_{r=t}.
\end{align}
According to \eq{hx}, $v(\3x,r)$ is the mean of 
$f$ over the sphere of radius $r$ centered at $\3x$,
hence $v(\3x, 0)=f(\3x, s)$.
Similarly,  \eq{hxs} states that $w(\3x,r)$ is the mean of 
$if_s$ over the same sphere, so $w(\3x, 0)=if_s(\3x, s)$.
We therefore need to show that $u$ solves the following
Cauchy problem: 
\begin{gather}\label{cauchy2}
u_{tt}(\3x,t)=\D u(\3x,t)\\
u(\3x,0)=v(\3x,0),\qq u_t(\3x,0)=w(\3x,0). \notag
\end{gather}
By the definition of $D_r$, the left equation in \eq{htilde2}
is
\begin{align}\label{first}
\frac {\sr{\p}}{2^k\G(k+\tfrac12)}
\,\pl_t\lp\Pl t\rp ^{\!\!k-1}
\!\! t^{2k-1}v(\3x,t)
=c_n\,\pl_t\lp\Pl t\rp ^{\!\! k-1} \!\! t^{2k-1}v(\3x,t),
\end{align}
where
\begin{align*}
c_n=\frac1{1\cdot3\cdots (2k-1)=1\cdot 3\cdots (n-2)}\,.
\end{align*}
The right equation in \eq{htilde2} is
\begin{align}\label{second}
c_n\lp\Pl r\rp ^{\!\! k}\lb\frac{r^2-t^2}{2k}\,
 r^{2k-1} w(\3x, r)\rb_{r=t}.
\end{align}
Letting $\x=r^2/2$,   a straightforward computation 
shows that for any function $G(\x)$,
\begin{align*}
\lp\Pl r\rp^{\!\! k}\lb(r^2-t^2) G\0\x\rb
&=\pl_\x^k\lb (2\x-t^2)G\0\x \rb
=2k\pl_\x^{k-1}G\0\x+(2\x-t^2)\,\pl_\x^kG\0\x\\
&=2k\lp\Pl r\rp^{\!\! k-1}\!\!G
+(r^2-t^2)\lp\Pl r\rp^{\!\! k}\!\! G.
\end{align*}
Thus \eq{second} becomes  
\begin{align}\label{second2}
c_n\lp\Pl r\rp^{\!\!k-1}\!\!
\lb r^{2k-1}w(\3x,r)\rb_{r=t}
=c_n\lp\Pl t\rp^{\!\!k-1}\!\!  t^{2k-1}w(\3x,t).
\end{align} 
The sum of \eq{first} and \eq{second2} is precisely the
solution $u(\3x,t)$ of the initial-value problem \eq{cauchy2}
with $n=2k+1$, as expressed in terms of \sl spherical means.
\rm  See John \ci{J55},  Courant and Hilbert 
\ci[pp.~699--703]{CH62} or Folland \ci[p.~170]{F95}.

That $u\in C^2(\4R^{n,1})$ follows because 
$v(\3x, r)\in C^{k+2}(\rr n)$ in $\3x$ and \eq{first}  contains
$k$ derivatives, while $w(\3x, r)\in C^{k+1}(\rr n)$ in $\3x$
and \eq{second2}  contains $k-1$ derivatives.  Finally,
uniqueness of the solution $u$ is a general property of the 
Cauchy problem. \sh2\qed  

As mentioned earlier, the above result can be extended to an
\sl even \rm number $n$ of space dimensions by applying the
recursion relation \eq{descent} between the singular
source distributions in $\cc n$ and $\4C^\nn$. In terms of
the solutions $\2f(\3x, s+it)$ to the Cauchy problem, this
amounts to using Hadamard's \sl method of descent \rm
\ci{H52, CH62}.  Consequently, \sl the same formula
\eq{htilde} gives the solution of the Cauchy problem for the
wave equation in $\4R^{n,1}$ for all values of $n\ge 2$. \rm 

The support properties of $\2\d$ now imply some 
important attributes of waves in $\4R^{n,1}$.
For $\/z=\/x+i\/y\in\4C^\nn$ with $\/y\ne\30$, recall that
\begin{align*}
\supp\2\d(\/x+i\/y)=
\begin{cases}
\5B(\/y) &\text{for odd  $n\ge 3$}\\
E\60(\/y) &\text{for all other $n\ge 2$}.\\
\end{cases}
\end{align*}
From \eq{htilde} we can therefore immediately draw the
following conclusions about \sl waves \rm (solutions of the
wave equation) $u(\3x, t)$ in $n$ space dimensions:

\begin{itemize}

\item 
For odd $n\ge 3$, $u(\3x,t)$ depends on the  values
of $u(\3x+\3v, 0)$ and   $u_t(\3x+\3v, 0)$ only in an
arbitrarily thin shell containing the sphere $|\3v|=t$. 
(We need a shell, rather than the sphere itself, 
because of the derivatives appearing in \eq{first} and
\eq{second2}.) This is the strong form of \sl Huygens' 
principle \rm \ci{BC87}, which states that  $u$
depends on the initial data only on the \sl light cone. \rm

\item 
For all other  $n\ge 2$, $u(\3x,t)$ depends on the  values
of $u(\3x+\3v, 0)$ and  $u_t(\3x+\3v, 0)$  in the \sl past
cone \rm $|\3v|\le t$. This is the \sl principle of
causality, \rm which states that no signal (information,
energy) can travel with speed greater than $c=1$. (If we
rescale time by $t\to ct$ with arbitrary $c>0$, then the
maximum propagation speed is $c$.)

\end{itemize}

We emphasize that  \sl   although holomorphy was needed in
the definition of $\2\d$, it is not necessary for the above
relation to the wave equation. \rm This  distinguishes our
results from all similar results in the literature of which I
am aware, where holomorphy of the Cauchy data is essential. 
See  Garabedian \ci[pp.~191--202]{G64} and Ryan 
\ci{R90, R90a, R96a}.

Theorem \ref{T:cauchy} and its counterpart for even $n$ can
be extended in various ways. 

\begin{itemize} 

\item
Clearly it is not necessary to
assume that $t>0$, since the support of $\2\d(\/x+i\/y)$ is
symmetric with respect to $\/y\to-\/y$.  When $t<0$,
$\2f(\3x, s+it)$ is a solution of the ``final-value problem'' in
terms of the Cauchy data at $t=0$.

\item
The Cauchy data $f(\/x)$ need not belong to
$C^{k+2}(\4R^{n+1})$. When $f$ is a distribution belonging,
say, to some Sobolev space \ci[Chapter 6]{F95}, then 
$\2f(\3x, s+it)$ is a \sl distributional \rm solution and the
derivatives in  \eq{first} and \eq{second2} must be
interpreted as distributional derivatives.

\item
To solve the \sl inhomogeneous \rm wave equation
\begin{gather*}
u_{tt}(\3x,t)-\D u(\3x,t)=j(\3x,t)\\
u(\3x,0)=v(\3x),\qq u_t(\3x,0)=w(\3x),
\end{gather*}
one can apply Duhamel's principle  to solutions of
the homogeneous equation \ci{F95}. This involves
integration of $\2\d(\/z')$ on the truncated solid  light cone 
with $0\le t'\le t$.  However, we will see that in the Clifford
setting, the current formalism leads directly to solutions of
inhomogeneous hyperbolic equations, where the
time-dependent source is determined by the given function
$f(\/x)$ in Euclidean spacetime.
 
\end{itemize}

\section{Extension to Clifford Analysis}
\lab{S: Clifford}

Clifford and quaternionic analyses \ci{BDS82, R96,
GS97,  R98} are  generalizations to
$\rr n$ of one-dimensional complex analysis that are proving
to be a unifying and very powerful tool in physics  \ci{H66,
KS96, MM98, O98, B99}. We now show that all the above
constructions generalize  naturally to this setting.  Let
$\clif n$ be the \sl complex \rm Clifford algebra generated
by elements $e_1,
\cdots, e_n$ satisfying the anticommutation relations 
\begin{align}\label{acr}
e_ke_l+e_l e_k=2\d_{k\,l}\qqq 1\le k,l\le n.
\end{align}
As a complex vector space, $\clif n$ has dimension $2^n$
with basis vectors
\begin{gather*}
e\6K\=e\6{k_1}\!\!\cdots e\6{k_p}\,, 
\qq K=\{k_1, \cdots, k_p\}, \qq  0\le p\le n,\qq
1\le k_1<\cdots <k_p\le n,
\end{gather*}
where the element labeled by the empty set $K=\varnothing$
($p=0$) is by definition $e_\varnothing=1$. Thus a general
vector  has the form $v=\sum_K  c\6K e\6K$, where
$c\6K\in\4C$ and the sum runs over the $2^n$ sets $K$ as
above.  The element $\sum_{k=1}^n z_k e_k\in\clif n$ is
identified with the vector $\3z\in\cc n$, and  by \eq{acr} the
products and squares of such vectors satisfy 
\begin{align}\label{z2}
\3z\3w+\3w\3z=2\sum_{k=1}^n z_k w_k\=2\,\3z\cdot\3w,
\qqq  \3z^2=\3z\cdot\3z\=\g(\3z)^2,
\end{align}
where $\g(\3z)$ is the complex Euclidean distance function. 
This connection will be our basis for generalizing holomorphic
potential theory to the Clifford setting. 

Clifford analysis is a noncommutative calculus dealing with
Clifford-valued functions 
\begin{align*}
f: \rr n\to\clif n, \qq
f(\3x)=\sum_K e_K f_K(\3x) \hb{where} f_K: \rr n\to\4C.
\end{align*}
The primary tool is the \sl Dirac operator \rm $\3D=\sum_k
e_k\pl_k$, which  is closely related to the exterior derivative 
\ci{AM78}  but in addition incorporates the underlying metric.
It acts on  Clifford-valued functions from either left or right 
by
\begin{gather}\label{dirac}
\3D f\=\sum_{k=1}^n e_k\,\ptl f{x_k}
\ne f\lv {{\3D}}\=\sum_{k=1}^n  \ptl f{x_k}e_k\\
\imp \3D^2f=f\lv {{\3D}}^2=\sum_K  e\6K\D f\6K=\D 
f.\label{D2}
\end{gather}  
Thus $\3D$ is a ``square root'' of the Laplacian $\D$ in $\rr n$.
It is an \sl elliptic \rm operator because the relations
\eq{acr} are based on the Euclidean metric
in $\rr n$.  In 1928, Dirac formulated a similar operator in
the Minkowskian  spacetime $\4R^{3,1}$, where it is  \sl 
hyperbolic \rm and its square is the wave operator. This
formed the basis for his relativistic wave equation of the
electron \ci{D58}, which had a revolutionary impact on
physics, including especially the dramatic prediction of
antimatter.   Mathematicians usually prefer Euclidean Dirac
operators because, among other things, they yield powerful
methods for solving boundary-value problems generalizing
those in one-dimensional complex analysis
\ci{GS97}. For this and similar reasons, most 
mathematical work on Dirac operators is restricted to
the elliptic case; see the discussion in  \ci[p.~1]{O98}. 
Consequently, rigorous analyses of Maxwell's equations by
Clifford methods  usually deal with static or
time-harmonic  fields  \ci{KS96, MM98}.  By assuming that the
boundary/Cauchy data is holomorphic, it is possible to arrive
at solutions of  hyperbolic Dirac equations through analytic
continuation \ci{R90, R90a, R96a},  generalizing the method 
employed by Garabedian for establishing a connection between
solutions of the Laplace and wave equations \ci{G64}.  As in
Section  \ref{S: Waves}, the present method  is \sl not \rm
 restricted by this assumption.  We will 
``Cliffordize'' the holomorphic potential theories in $\cc n$ 
and $\ccc n$, interpreted respectively as \sl complex space
\rm and \sl complex  spacetime. \rm The relation established
in Section
\ref{S: Waves} between Laplacians and  wave operators then
yields the desired connection between elliptic and hyperbolic
Dirac operators. 

The Clifford counterpart of the Newtonian potential $\f(\3x)$
is the \sl Cauchy kernel \rm $C: \rr n\to  \clif n$, defined as
the fundamental solution of $\3D$:
\begin{align*}
\3DC(\3x)=C(\3x)\lv{{\3D}}=\d(\3x), \qqq
\lim_{r\to\8}C(\3x)=0.
\end{align*}
Because $\3D^2=\D$, the
solution is easily expressed in terms of the Newtonian
potential:
\begin{align}\label{C(x)}
\f=\frac{r^{2-n}}{\o_n (2-n)}\,, \qqq
C(\3x)= \3D\f(\3x) =\frac{\3x}{\o_n \,r^n}.
\end{align}
The expression on  the right remains valid for $n=2$ if we
take $\f=(2\p)\inv\ln r$. Replacing $r(\3x)$ with $\g(\3z)$
immediately gives an extension to $\cc n$, where the point
source $\d$ becomes the extended source $\2\d$:
\begin{gather}
C(\3z)\=\3D\f(\3z)=\f(\3z)\lv{{\3D}}=\frac{\3z}{\o_n
\,\g^n}\,,
\qq\3z\in\cc n,\ n\ge 2\label{C(z)}\\
\imp \qq \3DC(\3z)=C(\3z)\lv{{\3D}}=\D\f(\3z)=\2\d(\3z).
\lab{DC}
\end{gather}
For all even $n\ge 2$,  $C(\3z)$ is holomorphic on the
complement of the null cone  $\5N$.  For
odd $n\ge3$, it inherits the branch cut from $\g$. Thus for all
$n\ge 2$, $C(\3z)$ is holomorphic on the complement of the
set
\begin{align*}
\5S_n=
\begin{cases}
\{\3x+i\3y\mid (\3x+i\3y)^2=0\} & \text{for even $n\ge 2$}\\
\{\3x+i\3y\in\4C^n\mid \3x\in E\60(\3y)\}
& \text{for odd $n\ge 3$}.\\
\end{cases}
\end{align*}

Given a Clifford-valued test function $f: \rr n\to \clif n$, we
define its extension to $\2f: \cc n\to\clif n$ exactly as
before:
\begin{align}\label{ftilde}
\2f(\3z)\=\int_\rr n \2\d(\3x'-\3z)\,f(\3x') \,d\3x',\imp
\lim_{\3y\to\30}\2f(\3x+i\3y)=f(\3x).
\end{align}
Recall that for $\2f$ to be defined,  $f$ must be $C^k$ with
$k=\lb \frac{n-1}2\rb$ if $n\ge 3$.  It can also be shown
\ci{K00} that  $f$ must be $C^1$ if $n=2$.  Inserting one of
the expressions \eq{DC} for $\2\d$ and integrating by parts
gives
\begin{align}\label{ftilde2}
\2f(\3z)=\int_\rr n C(\3x'-\3z)\lv{{\3D}}f(\3x') \,d\3x'
=-\int_\rr n C(\3x'-\3z) \3D f(\3x') \,d\3x',
\end{align}
where $\lv{{\3D}}$ denotes the left-acting Dirac opeator with
respect to $\3x'$. We will use this expression to  derive an
extended   version of the \sl Borel-Pompeiu formula. \rm
Let $M$ be a bounded domain in $\rr n$ with piecewise smooth
($C^1$) boundary  $\pl M$, and let
\begin{align*}
f\6M(\3x)=\c\6M(\3x)\,f(\3x), \hb{where}
\c\6M(\3x)=
\begin{cases}
1 &\text{ if }\3x\in M\\ 
0 &\text{ if }\3x\in \7M'\=\rr n\backslash \7M.\\
\end{cases}
\end{align*}
We do not need to define $f\6M$ on $\pl M$.
Taking the \sl distributional \rm derivatives of $f\6M$ gives
\begin{align}\label{DFM}
\3Df\6M(\3x)=(\3D\c\6M(\3x)) f(\3x)+\c\6M(\3x)\3Df(\3x).
\end{align}
We want to substitute this into \eq{ftilde2} to
obtain an expression for the extension $\2f\6M(\3z)$ of
$f\6M(\3x)$.  This will make sense if the singularities of
$\3Df\6M$ do not meet the singularities of $C(\3x-\3z)$. 
That will be the case if we assume that 
\begin{align}\label{zreg}
\3z&\notin \pl M+\5S_n
\=\{\3x_b+\3z\mid \3x_b\in\pl M,\ \3z\in\5S_n\},
\end{align}
and we refer to such points $\3z\in\cc n$ as \sl regular
\rm with respect to $\pl M$.  Note that when $\3z\in\rr n$,
this means simply that $\3z\notin \pl M$. For $\3z$ regular,
we may substitute \eq{DFM} into \eq{ftilde2} to obtain
\begin{align}\label{fMtilde}
\2f\6M(\3z)
&=-\int_\rr n C(\3x'-\3z)(\3D\c\6M(\3x')) f(\3x') \,d\3x'
-\int_\rr n C(\3x'-\3z) \c\6M(\3x')\3Df(\3x') \,d\3x'.
\end{align}
We claim that the first term can be written in classical
(non-distributional) form as
\begin{align}\label{classical}
\int_{\pl M} C(\3x'-\3z) \,\3n(\3x') f(\3x')\,d\s(\3x'),
\end{align}
where $\3n(\3x)$ is the outgoing unit normal at $\3x\in\pl
M$ and $d\s(\3x)$ is the area measure on $\pl M$ induced
from the volume measure $d\3x$. To see this, note that there
exists a differentiable function $\m(\3x)$ such that
\begin{align*}
M=\{\3x\mid \m(\3x)>0\}, \qq \pl M=\{\3x\mid \m(\3x)=0\},
\qq \nabla\m(\3x)=-|\nabla\m(\3x)|\,\3n(\3x).
\end{align*}
Then  $\c\6M$ can be expressed in terms of the Heaviside
step function $\q$ by $\c\6M(\3x)=\q(\m(\3x))$, hence
\begin{align*}
\3D\c\6M(\3x)=\d(\m(\3x)) \3D\m(\3x)
=\d(\m(\3x))\nabla\m(\3x)
=-\d(\m(\3x))\,|\nabla\m(\3x)|\,\3n(\3x).
\end{align*}
The connection to the expression \eq{classical} 
can now be made by using the implicit function theorem with
$\m(\3x)$ as one of the local coordinates.
The second term in \eq{fMtilde} reduces to the integral
over $M$, giving the following result.

\begin{thm}\label{T:ebp} 
{\bf (Extended Borel-Pompeiu formula)} 

 Let $\3z\in\cc n$ be  regular
with respect to $\pl M$. Then
\begin{align}\label{ebp}
\2f\6M(\3z)
&=\int_{\pl M} C(\3x'-\3z)\,\3n(\3x')f(\3x')\,d\s(\3x')
-\int_M C(\3x'-\3z)\, \3Df(\3x') \,d\3x'. \sh2\qed
\end{align}
\end{thm}

For $\3z\to\3x\in\rr n\backslash \pl M$, Equation
\eq{ftilde} applied to $f\6M$ reproduces $f\6M(\3x)$ since
$\2\d(\3x'-\3x)=\d(\3x'-\3x)$. Thus \eq{ebp} reduces to
the usual Borel-Pompeiu formula for Clifford-valued
functions \ci{GS97},
\begin{align}\label{bpf}
\int_{\pl M} C(\3x'-\3x)\,\3n(\3x')f(\3x')\,d\s(\3x')
-\int_M C(\3x'-\3x)\, \3Df(\3x') \,d\3x'=
\begin{cases}
f(\3x), & x\in M\\ 0,& \3x\in \7M'.\\
\end{cases}
\end{align}
We may interpret $f(\3x)$ as a field generated by the  source
function 
\begin{align}\label{j(x)}
j(\3x)\=\3Df(\3x).
\end{align}
Then \eq{bpf} solves the
boundary-value problem for \eq{j(x)}, expressing the field $f$
inside $M$ in terms of its sources in $M$ and its 
values on $\pl M$. (To investigate the limit of \eq{bpf} as
$\3x\to\pl M$, one also needs the \sl Plemelj-Sokhotzki
formula \rm
\ci{GS97}.)  In particular, $f$ is said to be
\sl left-monogenic \rm in $M$ if $j(\3x)=0$ in $M$. In that
case, the second term in \eq{bpf} vanishes and the
Borel-Pompeiu formula reduces to a multidimensional
generalization of Cauchy's integral formula, with
monogenicity replacing holomorphy.

To describe waves, such as a time-dependent
electromagnetic field, we ascend to $\ccc n$ as explained
earlier. In the notation of Section \ref{S: Waves}, let
\begin{gather*}
\3z=\3x+i\3y\in\cc n, 
\qq\/z=(\3z,  z_0)=\/x+i\/y\in\ccc n\\
\/y=(\30, t)\qq\/x=(\3x, s)
\imp\/z=(\3x, s+it). 
\end{gather*}
The generators of $\cliff n$ are $\{e_0, \cdots, e_n\}$,
satisfying relations identical to \eq{acr} but  with $0\le k,
l\le n$. The elliptic Dirac operators in $\rr n$ and $\rr\nn$ are
\begin{align*}
\3D=\sum_{k=1}^n e_k\pl_k, \qq
\/D\=\sum_{k=0}^n e_k\pl_k=\3D+e_0\pl_s\,, 
\qqq \pl_0\=\pl_s\,.
\end{align*}
We also define the \sl hyperbolic \rm (space-time) Dirac
operator by formally substituting $s\to it$ in $\/D$:
\begin{align*}
\rt D\=\3D-ie_0 \pl_t\,.
\end{align*}
Then
\begin{align*}
\/D^2=\D+\pl_s^2 \hb{and} \rt D^2=\D-\pl_t^2=\square,
\end{align*}
where $\D$ is the spatial Laplacian in $\rr n$ and
$\square$ is the wave operator in $\4R^{n, 1}$.

\begin{thm}\label{T: jtilde}
{\bf (Inhomogeneous Dirac/Maxwell Equation)}
For $n=2k+1\ge 3$, let $f: \rrr n\to\clif\nn$ be a 
$C^{k+2}$ function and define the functions
 $j: \rr\nn\to\clif \nn$ and $\2j: \cc\nn\to\clif \nn$ by
\begin{align}\label{jtilde}
\/D f(\3x, s)=j(\3x, s),\qqq 
\rt D \2f(\3x, s+it)=\2j(\3x, s+it).
\end{align}
Then $\2j(\3x, s+it)$ is  a $C^1$ solution of
\begin{align}\label{Djtilde}
\rt D \2j(\3x, s+it)=0,
\end{align}
and the system $\{\2f, \2j\}$ satisfies the initial
conditions
\begin{align}\label{MDic}
\2f(\3x, s)=f(\3x, s), \qqq \2j(\3x, s)=j(\3x, s).
\end{align}

\end{thm}

\sl Proof:  \rm This follows directly from Theorem
\ref{T:cauchy}. Since $\2f(\3x, s+it)$ is a
$C^2$ solution of the homogeneous wave equation,     
$\2j(\3x, s+it)$  is $C^1$ and 
\begin{align*}
\rt D \2j=\rt D^2\2f=\square \2f=0.
\end{align*}
From the initial conditions on $\2f$, we have
$\2f(\3x, s)=f(\3x, s)$ and
\begin{align*}
\lim_{t\to0}\2j(\3x, s+it)&=\3D\2f(\3x, s)
 -ie_0\lim_{t\to0}\pl_t\2f(\3x, s+it)\\
&=\3Df(\3x, s)+e_0 \pl_s f(\3x, s)=\/Df(\3x, s)=j(\3x, s).
\sh2\qed
\end{align*} 
For $n=3$,  the right side of \eq{jtilde} is precisely the
Clifford form of the \sl inhomogeneous, time-dependent
Maxwell equations \rm \ci{H66, O98, B99}, where
\begin{align}\label{F}
\2f(\3x, s+it)=\sum_{0=\m<\n}^3 e_{\m\n} F_{\m\n}(\3x, t)
\=F(\3x, t)
\end{align}
is the electromagnetic field (a bivector),
\begin{align}\label{J}
\2j(\3x, s+it)=\sum_{\m=0}^3 e_\m J_\m(\3x,t)\=J(\3x,t)
\end{align}
is the charge-current density (a four-vector), and the
Euclidean time $s$ is a free parameter. As with our
distance function, the Minkowskian metric appears in \eq{F}
and \eq{J} when the time components $J_0$  and $F_{0k}$ 
($k=1,2,3$) are imaginary  and the spatial components are
real.

Again, we did not need to assume that $\2f$ or $\2j$ are
holomorphic in $s+it$.  Note that although $\2j(\3x,
s+it)$ is a time-evolved extension of $j(\3x, s)$, it can be
shown \ci{K00} that it is \sl not \rm the one obtained by
convolving with  $\2\d$. Instead, the time dependence of the
charge-current density is governed by  \eq{Djtilde},  the
scalar part of which is the \sl continuity equation \rm
implying conservation of the total charge.

In  \eq{jtilde} and \eq{Djtilde}, we have taken
$M=\rr\nn$ for simplicity, giving a pure initial-value problem.
These equations may also be formulated  in a bounded region
$M\subset \rr\nn$ by using the extended Borel-Pompeiu
formula for $\2f\6M(\3x, s+it)$. This suggests that Theorem
\ref{T:ebp} may be used to solve  \sl mixed initial/boundary
value problems, \rm provided one is careful about dealing
with the singular points $(\3x, s+it)\in\pl M+\5S_\nn\,$.
These are in the  domain of influence  of the boundary;
that is, they  can be reached at time $t$ by signals
originating from the boundary $\pl M$ at time $t=0$. 
However, this still does not give a mechanism for 
boundary effects initiated at $t>0$ (such as reflections) to
influence the solution. This and related questions will be
treated elsewhere.

\sv3

\cl{\bf Acknowledgements}

This work was supported by AFOSR Contract 
\# F49620-98-C-0013.  It is a pleasure to thank Arje
Nachman for encouragement and Paul Garabedian, Sigurdur
Helgason, David Jerison, John Ryan and Frank Sommen for
stimulating and enjoyable conversations.

\VE

\end{document}